\documentclass[10pt,twoside,letter,amsmath,amssymb,aps,showkeys,showpacs]{revtex4}
\usepackage{amsfonts}
\usepackage{graphics}
\usepackage{epsfig}
\usepackage{color}

\begin{document}
\large

\thispagestyle{myheadings}

\title{
Quadratic electroweak corrections for polarized M{\o}ller scattering
}

\author{Aleksandrs Aleksejevs}
\email{aaleksejevs@swgc.mun.ca}
\affiliation{Memorial University, Corner Brook, Canada}

\author{Svetlana Barkanova}
\email{svetlana.barkanova@acadiau.ca}
\affiliation{Acadia University, Wolfville, Canada}

\author{Yury Kolomensky}
\email{yury@physics.berkeley.edu}
\affiliation{University of California, Berkeley, USA}

\author{Eduard Kuraev}
\email{kuraev@theor.jinr.ru}
\affiliation{Joint Institute for Nuclear Research, Dubna, Russia}

\author{Vladimir Zykunov}
\email{vladimir.zykunov@cern.ch}
\affiliation{Belarussian State University of Transport, Gomel, Belarus}

\begin{abstract}
The paper discusses the two-loop (NNLO) electroweak radiative corrections to the parity
violating $e^-e^- \rightarrow e^-e^- (\gamma)(\gamma\gamma)$
scattering asymmetry induced by squaring one-loop diagrams.
The calculations are relevant for the ultra-precise 11 GeV MOLLER
experiment planned at Jefferson Laboratory and experiments at future high-energy colliders.
The imaginary parts of the amplitudes are taken into consideration consistently
in both the infrared-finite and divergent terms.
The size of the obtained partial correction is significant, which indicates a need
for a complete study of the two-loop electroweak radiative corrections in order
to meet the precision goals of future experiments.
\end{abstract}


\maketitle

\section{Introduction}

Polarized M{\o}ller scattering has been a well-studied low-energy reaction for close
to eight decades now \cite{M1932}, but has attracted especially active interest from both
experimental and theoretical communities due to the recent rapid progress in measuring spin-dependent observables.
Since the nineties the interaction has allowed the high-precision determination of the electron-beam
polarization at SLC \cite{11}, SLAC \cite{12} \cite{13}, JLab \cite{14} and MIT-Bates \cite{15}.
A M{\o}ller polarimeter may also be useful in future experiments planned at the ILC \cite{18}.
In addition, polarized  M{\o}ller scattering can be an excellent tool for measuring parity-violating (PV)
weak interaction asymmetries \cite{DM1979}.

The first observation of Parity Violation in M{\o}ller scattering was made by the E-158 experiment
at SLAC \cite{2}, which studied M{\o}ller scattering of 45- to 48-GeV polarized electrons on the
unpolarized electrons in a hydrogen target. Its result at low $Q^2$ = 0.026 $\mbox{GeV}^2$,
$A_{PV} = (1.31 \pm 0.14\ \mbox{(stat.)} \pm 0.10\ \mbox{(syst.)}) \times 10^{-7}$ \cite{E158} allowed one
of the most important
parameters in the Standard Model -- the sine of the  Weinberg angle -- to be determined
with an accuracy of 0.5\% ($\sin^2 \theta_W$ = 0.2403  $\pm$ 0.0013 in the $ \overline{MS}$ scheme).
A very promising experiment measuring the e-p scattering asymmetry currently running at Jefferson Lab,
Qweak \cite{QWeak},  aims to determine $\sin^2 \theta_W $ with relative precision of 0.3\%.
The next-generation experiment to study e-e scattering -- MOLLER,  planned at JLab following the 11 GeV upgrade --
will offer a new level of sensitivity and measure the parity-violating asymmetry in the scattering
of longitudinally polarized electrons off an unpolarized target to a precision of 0.73 ppb.
That would allow a determination of the weak mixing angle with an uncertainty of
$\pm 0.00026 \ \mbox{(stat.)} \pm 0.00013 \ \mbox{(syst.)}$ \cite{JLab12}, or about 0.1\%, an
improvement of a factor of five in fractional precision when compared with the E-158 measurement.

Since M{\o}ller scattering  is a very clean process with a well-known
initial energy and low backgrounds, any inconsistency with the Standard Model will signal new physics. M{\o}ller scattering experiments can provide indirect access to physics at multi-TeV scales and play an important complementary role to the LHC research program \cite{1}.

Obviously, before we can extract reliable information from the experimental data, it is necessary
to take into account higher order effects of electroweak theory, i.e. electroweak radiative corrections (EWC). The inclusion of EWC is an indispensable part of any modern experiment, but will be of the paramount importance for the
ultra-precise measurement of the weak mixing angle via 11 GeV M{\o}ller scattering planned at JLab. In general, from the theory point of view, the interpretability of e-e scattering
is exceptionally good. However, to match the precision of MOLLER experiment,  theoretical predictions for the
PV e-e scattering asymmetry must include not only full treatment of one-loop radiative corrections (NLO) but also leading two-loop corrections (NNLO).

A significant theoretical effort has been dedicated to one-loop radiative corrections already. A short review of the literature to date on that topic is done in \cite{ABIZ-prd}. In \cite{ABIZ-prd}, we  also calculated a full gauge-invariant set of the one-loop EWC
both numerically with no simplifications using FeynArts \cite{int3}, FormCalc \cite{Hahn}, LoopTools \cite{Hahn} and Form \cite{int7} as the base languages and by hand in a compact form analytically free from nonphysical parameters. 
The total correction was found to be close to $-70$\%, and we found no significant theoretical uncertainty coming from the largest possible source, the hadronic contributions to the vacuum polarization.
The dependence on other uncertain input parameters, like the mass of the Higgs boson, was below 0.1\%.

It is possible that a much  larger theoretical uncertainty in the prediction for the asymmetry may come
from two-loop corrections.
Paper \cite{Petr2003} argued that the higher order corrections are suppressed by a factor of either about 0.1\% or 5\%
(depending on a type of corrections) relative to the one-loop result.  However, since the one-loop weak corrections
for  M{\o}ller scattering are so large  and since the 11 GeV MOLLER experiment is striving for such unprecedented precision, we believe it is now worth looking into evaluating two-loop weak corrections.

One way to find some indication
of the size of higher-order contributions is to compare results that are expressed in
terms of quantities related to different renormalization schemes.
In \cite{arx-2}, we provided a tuned comparison between the result obtained with different renormalization conditions, first within one scheme then between two schemes.
Our calculations in the on-shell and CDR schemes show a difference of about 11\%,
which is comparable with the difference of 10\% between
$\rm \overline{MS}$ \cite{Czar1996} and the on-shell scheme  \cite{Petr2003}.
It is also worth noting that although two-loop corrections to the cross section may seem to be small,  it is much harder to estimate their scale and behaviour for such a complicated observable as the
parity-violating asymmetry to be measured by MOLLER experiment.

The two-loop EWC to the Born ($\sim M_0M_0^+$) cross section can be divided into two classes:
the Q-part induced by quadratic one-loop amplitudes ($ \sim M_1M_1^+$),
and the T-part corresponding to the interference of the Born and two-loop diagrams ($ \sim 2 \mbox{\rm Re} M_{0} M_{2-loop}^+$).
The goal of this paper is to calculate the Q-part exactly.
We show that the Q-part is much higher than the planned experimental uncertainty of MOLLER,
which means that the two-loop EWC may be larger that previously thought. The large size of
the Q-part demands a detailed and consistent consideration of the T-part, and that will be the next task of our group.

\section{General notations and matrix elements}

\begin{figure}
\vspace{10mm}
\begin{tabular}{cc}
\begin{picture}(60,60)
\put(-150,-60){
\epsfxsize=7cm
\epsfysize=7cm
\epsfbox{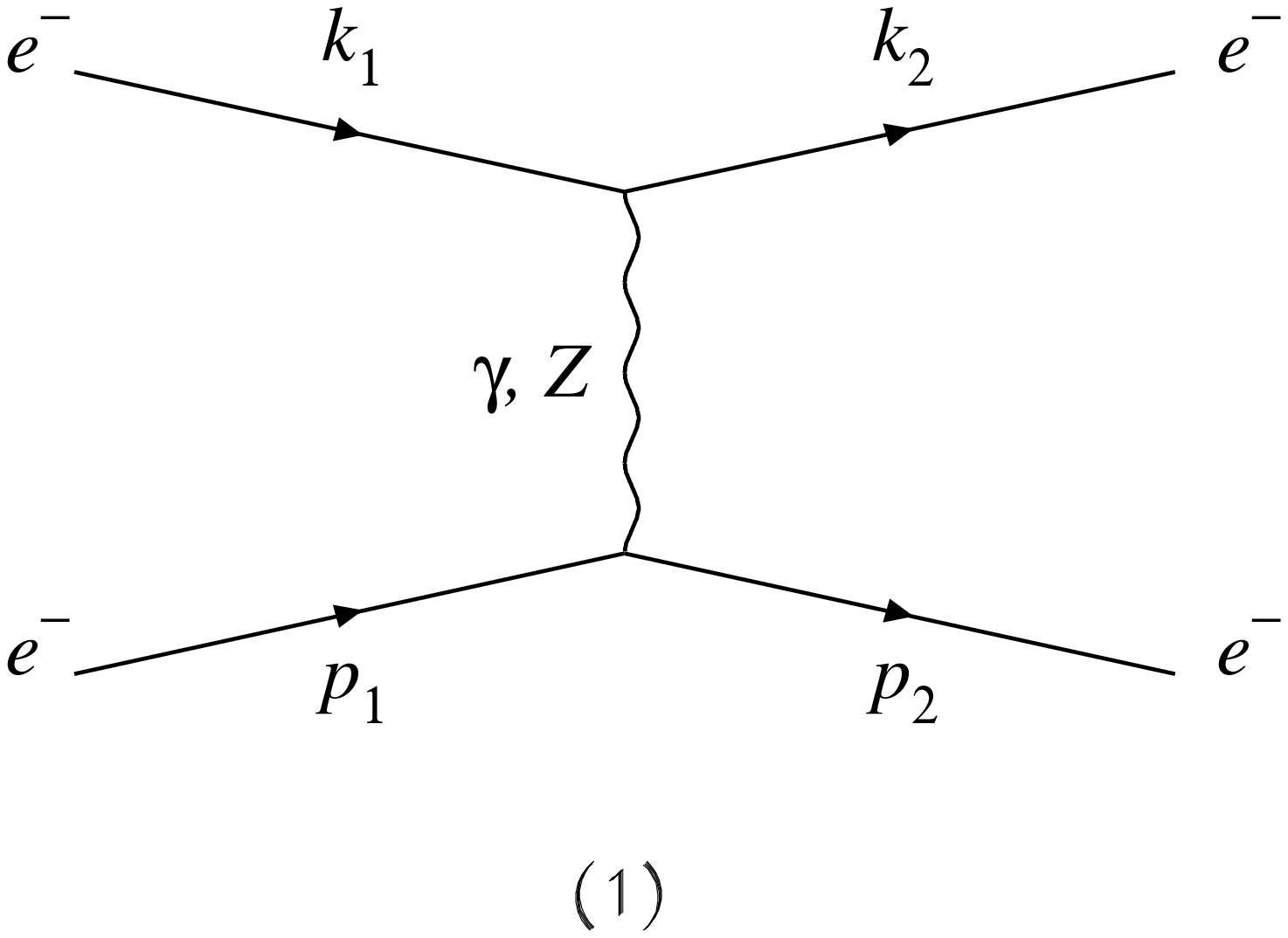} }
\end{picture}
&
\begin{picture}(60,100)
\put(-20,-60){
\epsfxsize=7cm
\epsfysize=7cm
\epsfbox{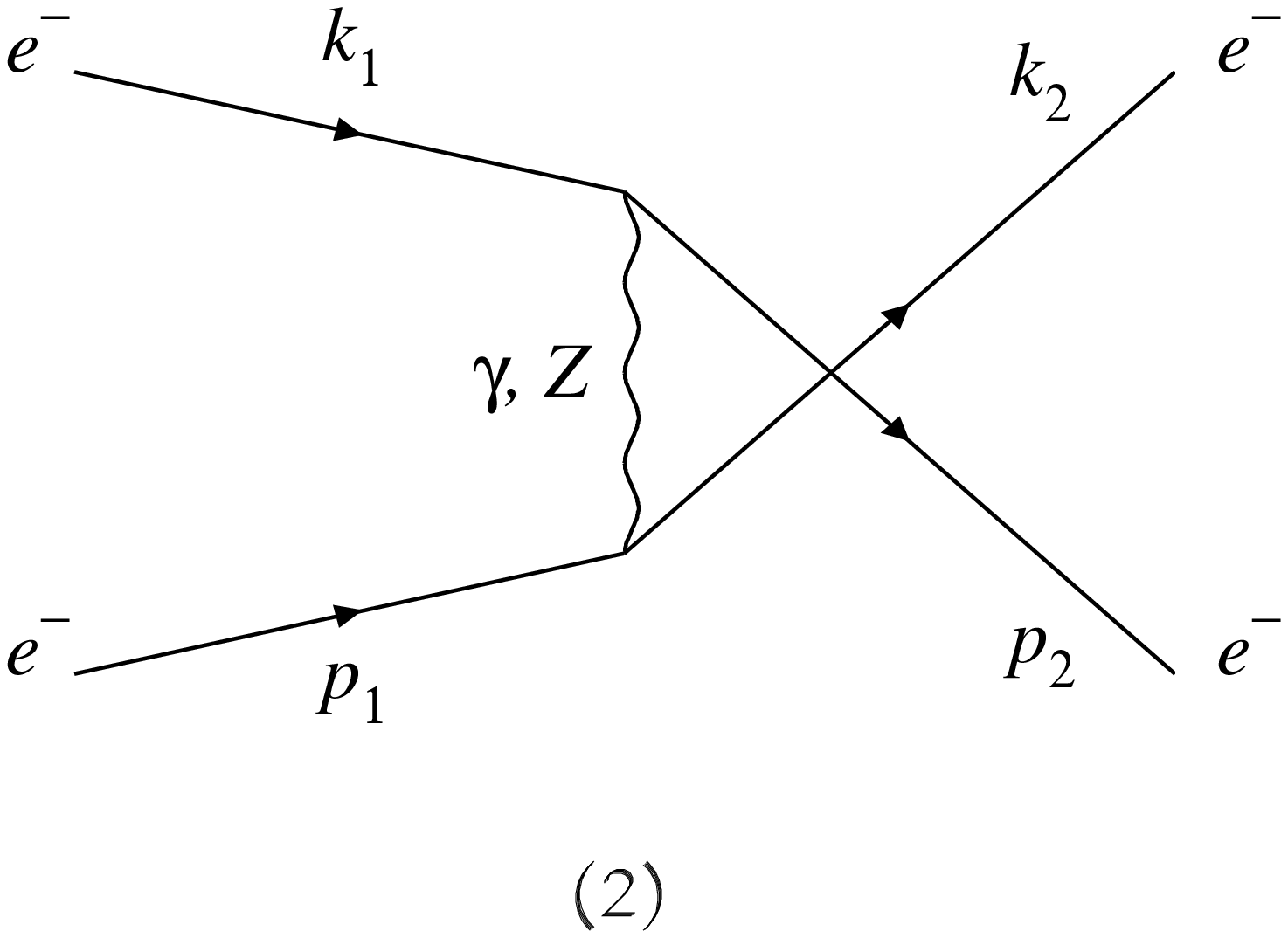} }
\end{picture}
\end{tabular}
\vspace{10mm}
\caption{\protect\it
Diagrams describing nonradiative M{\o}ller scattering in the (1) t- and (2) u-channels.}
\label{born}
\vspace{5mm}
\end{figure}

Let us start by writing the cross section of polarized M{\o}ller scattering
with the Born kinematics shown in Fig. \ref{born},
\begin{equation}
e^-(k_1)+e^-(p_1) \rightarrow e^-(k_2)+e^-(p_2),
\label{0}
\end{equation}
such that, with the appropriate accuracy for the present paper, we find:
\begin{equation}
\sigma = \frac{\pi^3}{2s} |M_0+M_1|^2 = \frac{\pi^3}{2s} (M_0M_0^+ + 2 {\rm Re} M_1M_0^+ + M_1M_1^+).
\label{01}
\end{equation}
Here,
$\sigma \equiv {d\sigma}/{d \cos \theta}$,
$\theta$  is the scattering angle of the detected electron
with 4-momentum $k_2$ in the center-of-mass system of the initial electrons. The
 4-momenta of initial ($k_1$ and $p_1$) and final
($k_2$ and $p_2$) electrons generate a standard
set of Mandelstam variables:
\begin{equation}
s=(k_1+p_1)^2,\ t=(k_1-k_2)^2,\ u=(k_2-p_1)^2.
\label{stu}
\end{equation}
It should also be noted that the electron mass $m$ is
disregarded wherever possible, in particular if
$m^2 \ll s,-t,-u$.

$M_0$ and $M_1$ are the Born (${\cal O}(\alpha)$)
and one-loop (${\cal O}(\alpha^2)$) amplitudes (matrix elements), respectively.
Let us describe the structure of $M_0$:
\begin{eqnarray}
M_0 = M_{0,t}-M_{0,u},\
M_{0,u}=M_{0,t}(k_2 \leftrightarrow p_2),\
M_{0,t} = \sum_{j=\gamma,Z} M_{t}^j,\
M_{t}^j = i\frac{\alpha}{\pi} I_\mu^j  D^{jt} J^{\mu,j},
\end{eqnarray}
where the $t$-channel upper and lower electron leg currents are
\begin{eqnarray}
I_\mu^j = \bar u(k_2)\gamma_\mu (v^j-a^j\gamma_5) u(k_1),\
J_\mu^j = \bar u(p_2)\gamma_\mu (v^j-a^j\gamma_5) u(p_1).
\end{eqnarray}

The squared Born amplitude $M_0$  forms the Born cross section:
\begin{equation}
\sigma_0 =\frac{\pi^3}{2s} M_0M_0^+ = \frac{\pi \alpha^2}{s}
\sum_{i,j=\gamma,Z} [\lambda_-^{i,j}(u^2D^{it}D^{jt}+t^2D^{iu}D^{ju})
   + \lambda_+^{i,j}s^2(D^{it}+D^{iu})(D^{jt}+D^{ju})].
\label{cs0}
\end{equation}

A handy structure to use in the present study is
\begin{equation}
D^{ir}=\frac{1}{r-m_i^2}\ \ (i=\gamma,Z;\ r=t,u),
\label{structure}
\end{equation}
which depends on the $Z$-boson mass $m_Z$ or
on the photon mass $m_\gamma \equiv \lambda $. The photon mass  is set to zero
everywhere with the exception of specially-indicated
cases where the photon mass is taken to be an
infinitesimal parameter that regularizes the infrared
divergence (IRD). Another set of useful functions is
\begin{eqnarray}
{\lambda_{\pm}}^{i,k} =
     {\lambda_1}_B^{i,k}{\lambda_1}_T^{i,k} \pm {\lambda_2}_B^{i,k}{\lambda_2}_T^{i,k},
\label{b10}
\end{eqnarray}
These are combinations of coupling constants
and $p_{B(T)}$, where $p_{B(T)}$ are the degrees of polarization
of electrons with 4-momentum $k_1$ ($p_1$). More specifically,
\begin{eqnarray}
 {\lambda_1}_{B(T)}^{i,j} = \lambda_V^{i,j} -p_{B(T)} \lambda_A^{i,j},\
   {\lambda_2}_{B(T)}^{i,j} = \lambda_A^{i,j} -p_{B(T)} \lambda_V^{i,j},
\nonumber
\end{eqnarray}
\begin{equation}
 \lambda_V^{i,j}=v^iv^j + a^ia^j,\
   \lambda_A^{i,j}=v^ia^j + a^iv^j,
\end{equation}
where
\begin{equation}
 v^{\gamma}=1,\ a^{\gamma}=0,\
 v^Z=(I_e^3+2s_{W}^2)/(2s_{W}
c_{W}),
\ a^Z=I_e^3/(2s_{W}c_{W}).
\end{equation}
The subscripts $L$ and $R$ on the cross sections correspond
to $p_{B(T)}$ = $-1$ and $p_{B(T)}$ = $+1$, where the first subscript
indicates the degree of polarization for the
4-momentum $k_1$ and the second one indicates the
degree of polarization for the 4-momentum $p_1$.
Let us recall that $ I_e^3=-1/2 $ and
$s_{W}\  (c_{W})$
is the sine (cosine) of the Weinberg angle expressed in terms of the $Z$- and $W$-boson
masses according to the rules of the Standard Model:
\begin{equation}
c_{W}=m_{W}/m_{Z},\
s_{W}=\sqrt{1-c_{W}^2}.
\end{equation}

\begin{figure}
\vspace{15mm}
\begin{tabular}{ccc}
\begin{picture}(60,60)
\put(-120,0){
\epsfxsize=5cm
\epsfysize=5cm
\epsfbox{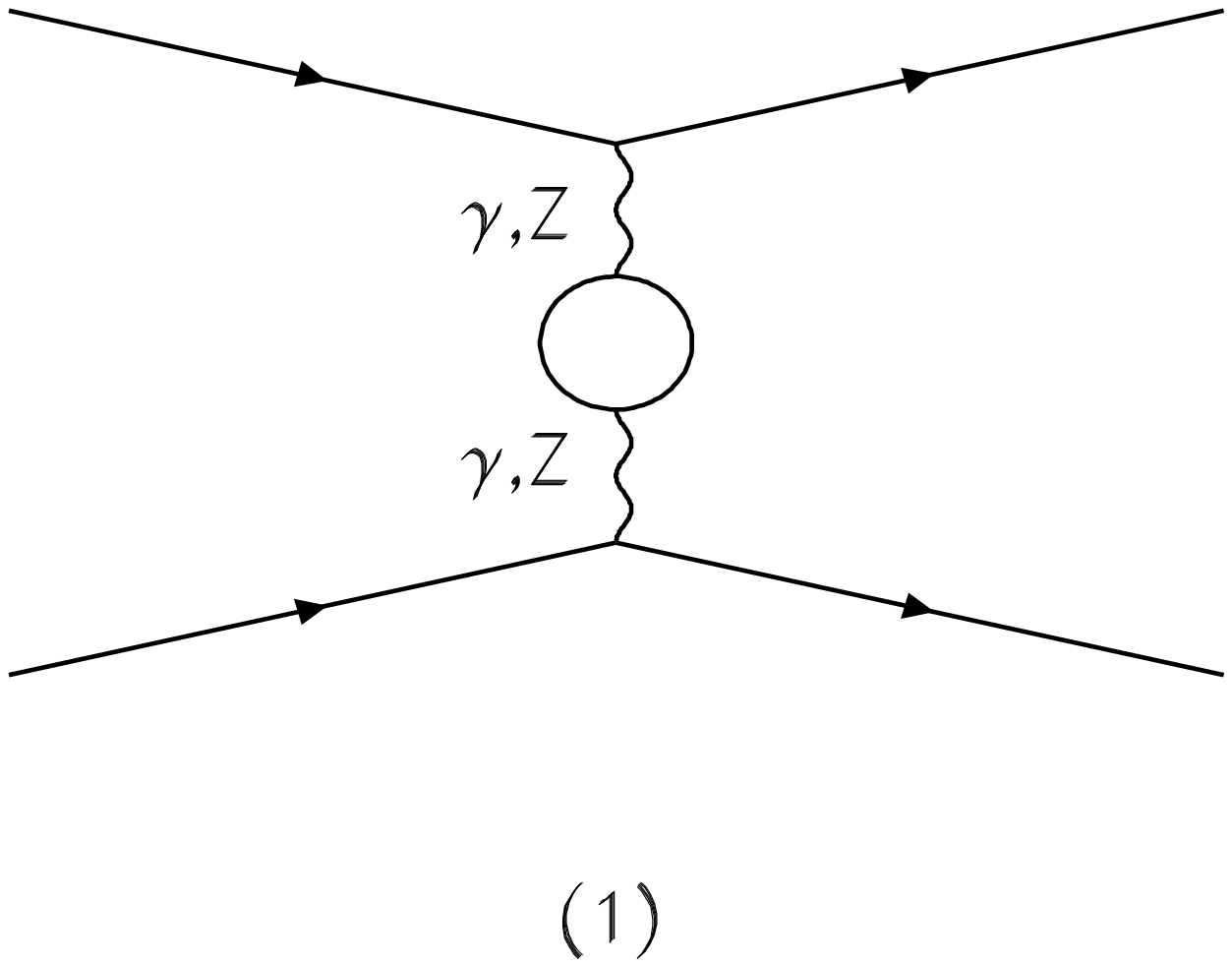} }
\end{picture}
&
\begin{picture}(60,100)
\put(-70,0){
\epsfxsize=5cm
\epsfysize=5cm
\epsfbox{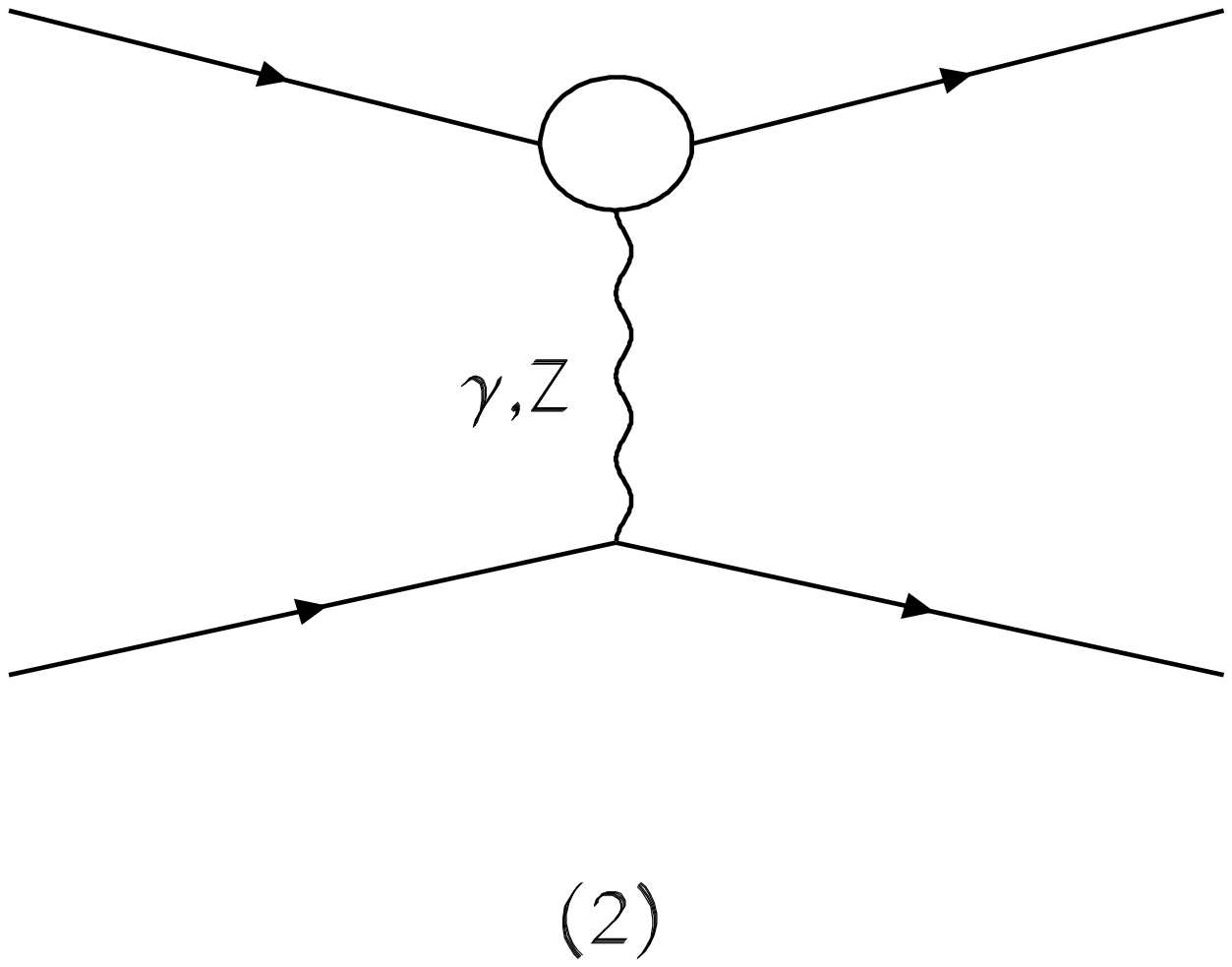} }
\end{picture}
&
\begin{picture}(60,100)
\put(-20,0){
\epsfxsize=5cm
\epsfysize=5cm
\epsfbox{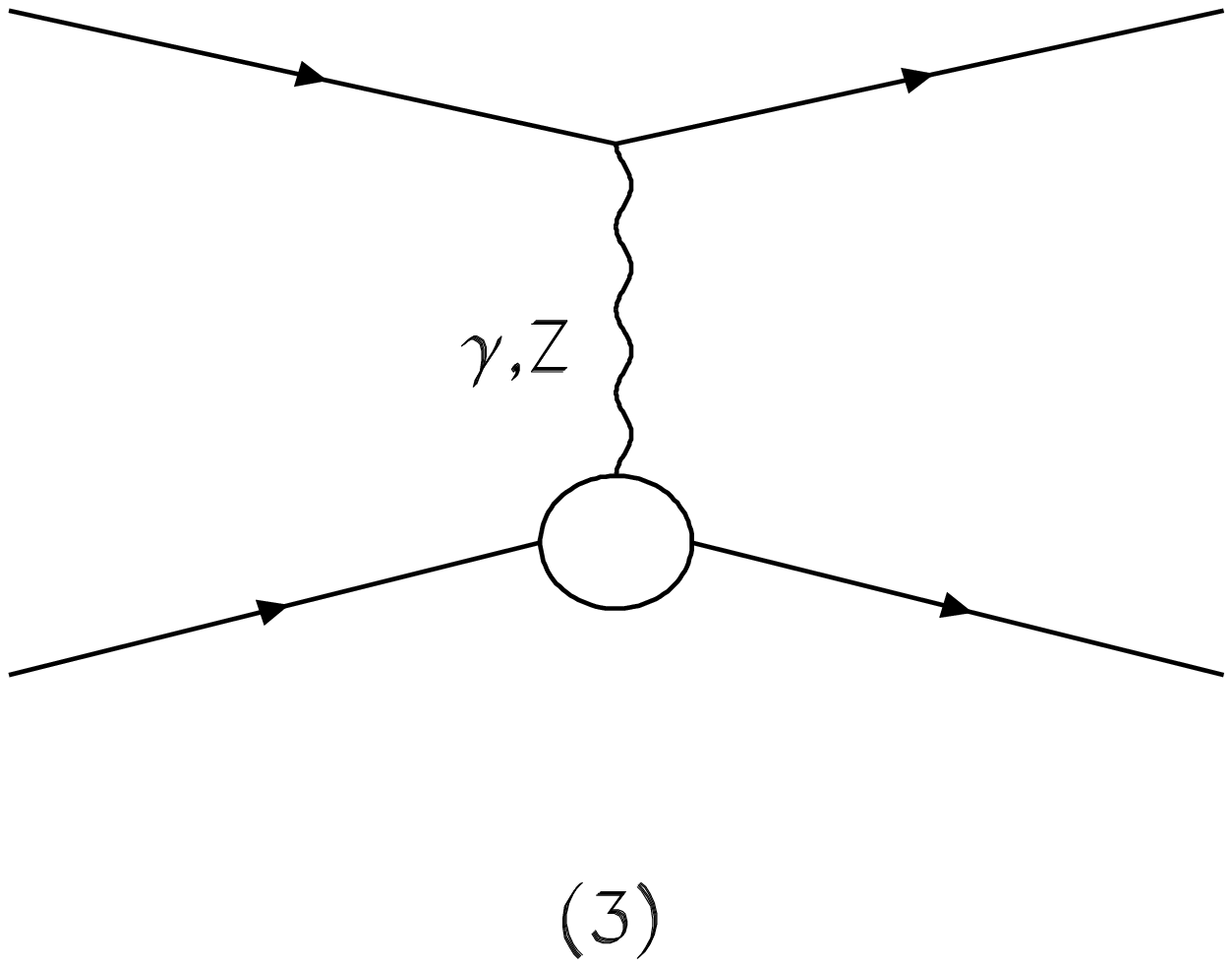} }
\end{picture}
\\[3mm]
\begin{picture}(60,60)
\put(-60,0){
\epsfxsize=5cm
\epsfysize=5cm
\epsfbox{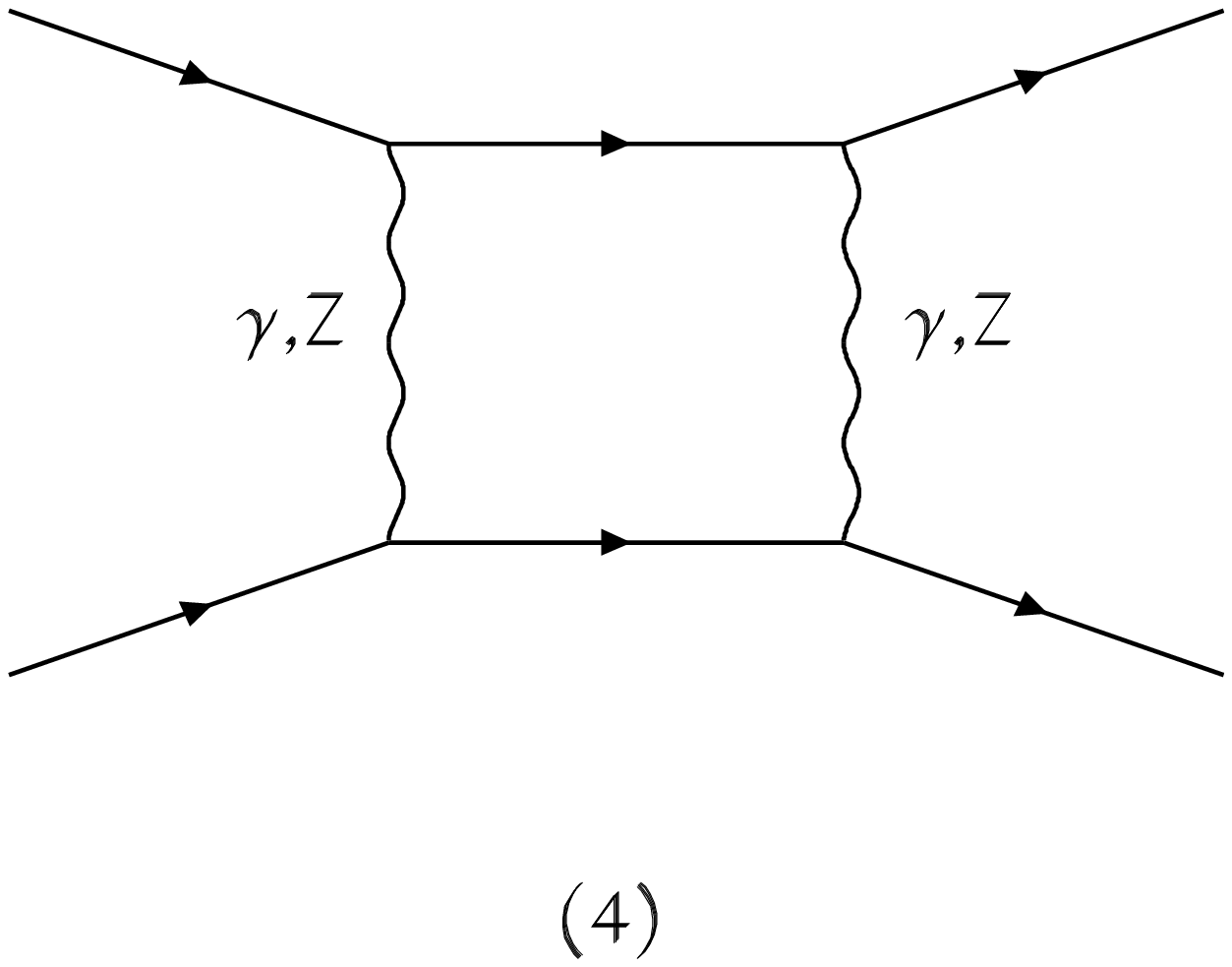} }
\end{picture}
&
\begin{picture}(60,100)
\put(-10,0){
\epsfxsize=5cm
\epsfysize=5cm
\epsfbox{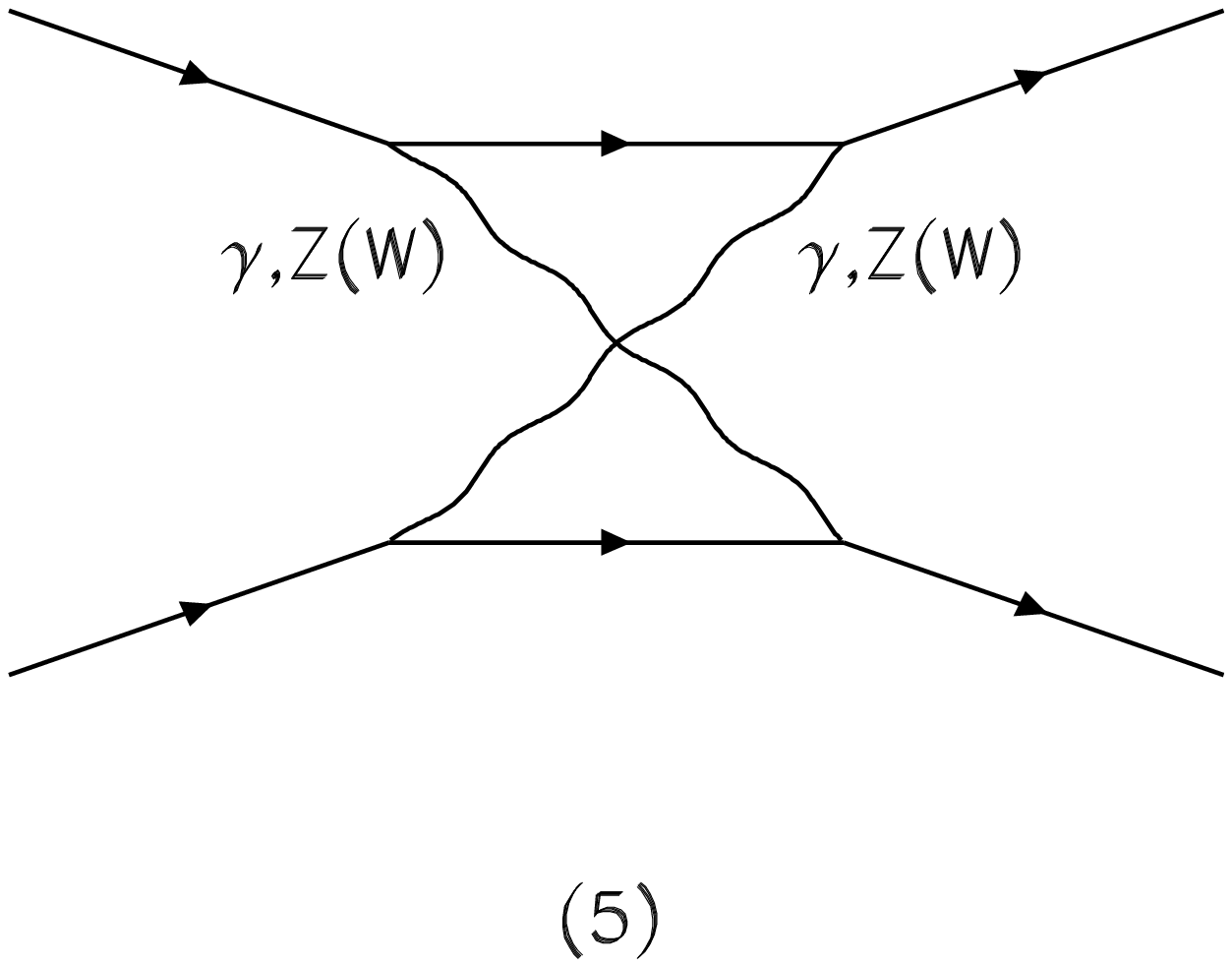} }
\end{picture}
\end{tabular}
\vspace*{-20mm}
\caption{\protect\it
One-loop t-channel diagrams for the M{\o}ller process.
The circles represent the contributions of self-energies
and vertex functions.
The u-channel diagrams are obtained via the interchange $k_2 \leftrightarrow p_2$. }
\label{2f}
\vspace{5mm}
\end{figure}

We can present the one-loop amplitude $M_1$ as a sum of boson self-energy (BSE),
vertex (Ver) and box diagrams:
\begin{eqnarray}
 &&
M_1 = M_{1,t}-M_{1,u},\
M_{1,u}=M_{1,t}(k_2 \leftrightarrow p_2),\
 \nonumber \\
 &&
M_{1,t}=M_{{\rm BSE},t}+M_{{\rm Ver},t}+M_{{\rm Box},t}.
\end{eqnarray}
We use the on-shell renormalization scheme from \cite{BSH86, Denner},
so there are no contributions from the electron self-energies.
The question of the dependence of EWC on
renormalization schemes and renormalization conditions (within the same scheme) was addressed in our earlier paper \cite{arx-2}.

The infrared-finite BSE term can easily be expressed as:
\begin{equation}
M_{{\rm BSE},t}= i\frac{\alpha}{\pi}  \sum_{i,j=\gamma,Z} I_\mu^i  D^{ijt}_S J^{\mu,j},
\end{equation}
with
\begin{equation}
D_S^{ijr}=-D^{ir} {\hat{\Sigma}}_T^{ij}(r) D^{jr},
\label{D_S}
\end{equation}
where ${\hat{\Sigma}}_T^{ij}(r)$
is the transverse part of the renormalized
photon, $Z$-boson and $\gamma Z$ self-energies.
The longitudinal parts of the boson self-energy make contributions
that are proportional to  $m^2/r$;
therefore they are very small and are not considered here.

In order to get the
electron vertex amplitude
(2nd and 3rd diagrams in Fig.~\ref{2f}),
we use the form factors $\delta F_{V,A}^{je}$ in the manner of paper \cite{BSH86},
replacing the coupling
constants $v^j,\ a^j$ with form factors
 $v^{\gamma (Z)} \rightarrow \delta F_V^{\gamma (Z) e}$,\
 $a^{\gamma (Z)} \rightarrow \delta F_A^{\gamma (Z) e}$.
Then,
\begin{equation}
M_{{\rm Ver},t} = \sum_{j=\gamma,Z} \Bigl( M_{j/B,t} + M_{j/H,t} \Bigr),\
M_{j/B,t} = i\frac{\alpha}{\pi} B_\mu^j  D^{jt} J^{\mu,j},\
M_{j/H,t} = i\frac{\alpha}{\pi} I_\mu^j  D^{jt} H^{\mu,j},
\end{equation}
where the electron currents with vertices look like
\begin{equation}
B_\mu^j = I_\mu^j (v^{j} \rightarrow \delta F_V^{j e},\ a^{j} \rightarrow \delta F_A^{j e}),\
H^{\mu,j} =J^{\mu,j} (v^{j} \rightarrow \delta F_V^{j e},\ a^{j} \rightarrow \delta F_A^{j e}).
\end{equation}
The infrared singularity is regularized by giving photon a small mass $\lambda$
and in the $t$-channel vertex amplitude can be extracted in the form:
\begin{equation}
M_{{\rm Ver},t}^\lambda =
-\frac{\alpha}{\pi} \Bigl( \log\frac{-t}{m^2} -1 \Bigr)  \log\frac{s}{\lambda^2} M_{0,t},
\end{equation}
where $e$ is the base of the natural logarithm.
The rest (infrared-finite) part of $t$-channel vertex amplitude has the simple form
The remaining (infrared-finite) part of the $t$-channel vertex amplitude has a simple form
convenient for analysis and coding:
\begin{equation}
M_{{\rm Ver},t}^f =
M_{{\rm Ver},t} - M_{{\rm Ver},t}^\lambda =
M_{{\rm Ver},t}(\lambda^2 \rightarrow s).
\end{equation}

The box term can be presented as a sum of all two-boson contributions:
\begin{equation}
M_{{\rm Box},t}= M_{\gamma\gamma,t} +M_{\gamma Z,t} +M_{Z\gamma,t} + M_{ZZ,t} +M_{WW,t}.
\end{equation}
We need to account for both direct and crossed $\gamma\gamma$, $\gamma Z$ and $ZZ$-boxes:
\begin{equation}
M_{ij,t}= M_{ij,t}^D + M_{ij,t}^C\ \ (i,j=\gamma,Z),
\end{equation}
with $M_{ij,t}^D$ and $M_{ij,t}^C$ given by exact expressions in 4-dimensional integral form (4-point functions) by
\begin{eqnarray}
M_{ij,t}^D =
-i \Bigl(\frac{\alpha}{\pi} \Bigr)^2\cdot \frac{i}{(2\pi)^2}
 &&
\int \frac{d^4k}{(k^2-2k_1k)(k^2+2p_1k)((q-k)^2-m_j^2)(k^2-m_i^2)} \times
 \nonumber \\
 &&
\times \bar u(k_2)\gamma^\mu (v^j-a^j\gamma_5)(\hat k_1-\hat k+m)\gamma^\nu (v^i-a^i\gamma_5) u(k_1) \times
 \nonumber \\
 &&
\times \bar u(p_2)\gamma_\mu (v^j-a^j\gamma_5)(\hat p_1+\hat k+m)\gamma_\nu (v^i-a^i\gamma_5 u(p_1),
\end{eqnarray}
\begin{eqnarray}
M_{ij,t}^C =
-i \Bigl(\frac{\alpha}{\pi} \Bigr)^2\cdot \frac{i}{(2\pi)^2}
 &&
\int \frac{d^4k}{(k^2-2k_1k)(k^2-2p_2k)((q-k)^2-m_j^2)(k^2-m_i^2)} \times
 \nonumber \\
 &&
\times \bar u(k_2)\gamma^\mu (v^j-a^j\gamma_5)(\hat k_1-\hat k+m)\gamma^\nu (v^i-a^i\gamma_5) u(k_1) \times
 \nonumber \\
 &&
\times \bar u(p_2)\gamma_\nu (v^i-a^i\gamma_5)(\hat p_2-\hat k+m)\gamma_\mu (v^j-a^j\gamma_5) u(p_1).
\label{cr}
\end{eqnarray}
Obviously, for $WW$-boxes we only need the crossed expression (\ref{cr}).

The infrared parts of the $\gamma\gamma$- and $\gamma Z$-boxes in the $t$-channel are similarly given by
\begin{eqnarray}
M_{\gamma\gamma (\gamma Z + Z \gamma),t}^\lambda =
 &&
-\frac{\alpha}{\pi} \Bigl( \frac{1}{2} \log\frac{-u}{s}\log\frac{-us}{\lambda^4} +\frac{\pi^2}{2}
+ i\pi \log\frac{s}{\lambda^2} \Bigr) M^{\gamma (Z)}_t.
\label{box-la}
\end{eqnarray}
Using asymptotic methods, we can significantly simplify the box amplitudes
containing at least one heavy boson (see, for example, \cite{ABIZ-prd}, where simplifications were done
on the cross-section level).
Then
\begin{eqnarray}
M_{\gamma Z,t}^f + && M_{Z\gamma,t}^f
=
\Bigl( M_{\gamma Z,t} +  M_{Z\gamma,t}\Bigr)
- \Bigl( M_{\gamma Z,t}^\lambda + M_{Z\gamma,t}^\lambda  \Bigr)
=
 - 2 i \Bigl(\frac{\alpha}{\pi} \Bigr)^2 \times
 \nonumber \\
 \times
\Biggl[
\Bigl(  && \frac{3}{2}+\log\frac{m_Z^2}{s} \Bigr)
 \bar u(k_2)\gamma^\mu (v^Z-a^Z\gamma_5)(-\gamma^\alpha)\gamma^\nu u(k_1) \cdot
 \bar u(p_2)\gamma_\mu (v^Z-a^Z\gamma_5) \gamma_\alpha \gamma_\nu u(p_1) +
 \nonumber \\
+
\Bigl( && \frac{3}{2}+\log\frac{m_Z^2}{-u} \Bigr)
 \bar u(k_2)\gamma^\mu (v^Z-a^Z\gamma_5)\gamma^\alpha\gamma^\nu u(k_1) \cdot
 \bar u(p_2)\gamma_\nu \gamma_\alpha \gamma_\mu  (v^Z-a^Z\gamma_5) u(p_1) \Biggr],
\end{eqnarray}
\begin{eqnarray}
M_{ZZ,t}
=  - i \Bigl(\frac{\alpha}{\pi} \Bigr)^2   \frac{1}{16 m_Z^2}  \Biggl[
 &&
 \bar u(k_2)\gamma^\mu (v^B-a^B\gamma_5)(-\gamma^\alpha)\gamma^\nu u(k_1) \cdot
 \bar u(p_2)\gamma_\mu (v^B-a^B\gamma_5) \gamma_\alpha \gamma_\nu u(p_1) +
 \nonumber \\
+
 &&
 \bar u(k_2)\gamma^\mu (v^B-a^B\gamma_5)\gamma^\alpha\gamma^\nu u(k_1) \cdot
 \bar u(p_2)\gamma_\nu \gamma_\alpha \gamma_\mu  (v^B-a^B\gamma_5) u(p_1) \Biggr],
\end{eqnarray}
\begin{eqnarray}
M_{WW,t}
=  - i \Bigl(\frac{\alpha}{\pi} \Bigr)^2   \frac{1}{16 m_W^2}  \Biggl[
 \bar u(k_2)\gamma^\mu (v^C-a^C\gamma_5)\gamma^\alpha\gamma^\nu u(k_1) \cdot
 \bar u(p_2)\gamma_\nu \gamma_\alpha \gamma_\mu  (v^C-a^C\gamma_5) u(p_1) \Biggr],\ \ \ \ \ \
\end{eqnarray}
with the coupling-constants combinations for $ZZ$- and $WW$-boxes
\begin{equation}
v^B={(v^Z)}^2+{(a^Z)}^2,\ a^B=2v^Za^Z,\ v^C=a^C=1/(4s_W^2).
\end{equation}

Now we are ready to present the one-loop complex amplitude as the sum of IR and IR-finite parts:
\begin{equation}
M_1 = M_1^\lambda + M_1^f,\
M_1^\lambda = \frac{\alpha}{\pi} \frac{1}{2} {\delta_1^{\lambda}} M_0,\
M_1^f =  M_{\rm BSE}+M_{\rm Ver}^f+M_{\rm Box}^f+M_a,
\label{mmm}
\end{equation}
where
\begin{equation}
\delta_1^{\lambda} = 4 B \log\frac{\lambda}{\sqrt{s}},
\end{equation}
and the complex value $B$ can be presented in form (see, for example, \cite{KuFa})
\begin{equation}
B= \log\frac{tu}{m^2s}-1 +i\pi.
\end{equation}
The amplitudes from the non-factorized part of the boxes are given by
\begin{equation}
{\rm Re} M_a=-\frac{\alpha}{2\pi} \Bigl[ (L_u^2+\pi^2)M_{0,t} - (L_t^2+\pi^2)M_{0,u} \Bigr].
\end{equation}
where $L_r=\log(-s/r)$.

\section{Extraction of infrared divergences}

Now we should make sure that the infrared divergences  are cancelled.
In a similar way as it was done for amplitudes, we present
the complex interference term $\hat \sigma_1$ and differential
cross section  $\sigma_Q$ as sums of $\lambda$-dependent
(IRD-terms) and $\lambda$-independent (infrared-finite) parts:
\begin{equation}
\hat \sigma_1 = \frac{\pi^3}{s}  M_1M_0^+ = \sigma^{\lambda}_1 + \sigma^{f}_1,\ \
\sigma_Q = \frac{\pi^3}{2s} M_1M_1^+   = \sigma^{\lambda}_Q + \sigma^{f}_Q.
\end{equation}
The one-loop cross section which we denote
$\sigma_1 ={\rm Re} \hat\sigma_1 $ was carefully evaluated with full control
of the uncertainties in paper \cite{ABIZ-prd}.
The term $\sigma_Q$ (see (\ref{01})) is called the Q-part of the two-loop EWC and
is the main subject of the present paper.

If we substitute the amplitudes derived in Section II to the left-hand-side of (\ref{01}),
and compare the result with the right-hand side of this equation, we will get the same
expression for $\sigma_1$ as given in \cite{ABIZ-prd}.
The simplest form for $\sigma^{\lambda}_1$ (see formula (42) of \cite{ABIZ-prd}) is then:
\begin{equation}
\sigma^{\lambda}_{1} = \frac{\alpha}{\pi} \delta_1^{\lambda} \sigma_0.
\end{equation}
The infrared-finite part $\sigma^{f}_1$ can be conveniently to presented via the relative
dimensionless correction:
\begin{equation}
\sigma^{f}_{1} = \frac{\alpha}{\pi}  \delta_1^f \sigma_0.
\end{equation}

After some transformations, the value $\sigma^{\lambda}_Q$ is given by
\begin{eqnarray}
\sigma^{\lambda}_Q =
\frac{\pi^3}{2s} {M_1^\lambda}^+ \bigl( M_1^\lambda + 2 M_1^f \bigr) =
\frac{1}{4} {\Bigl( \frac{\alpha}{\pi} \Bigr)}^2
{\rm Re} \Bigl[ {\delta_1^{\lambda}}^* (\delta_1^{\lambda}+2\delta_1^{f}) \Bigr] \sigma_0.
\label{simpl2}
\end{eqnarray}
Finally, the infrared-finite part $\sigma^{f}_Q$ expressed via the relative
dimensionless corrections has form
\begin{equation}
\sigma^{f}_Q =
\frac{\pi^3}{2s} M_1^f {M_1^f}^+ =
{\Bigl( \frac{\alpha}{\pi} \Bigr)}^2 \delta_Q^f \ \sigma_0.
\end{equation}

\section{Bremsstrahlung and cancellation of infrared divergences }

To evaluate the cross section induced by the emission of one soft photon with energy less then $\omega$,
we follow the methods of  \cite{HooftVeltman} (see also	\cite{KT1}).
Then this cross section can be expressed as:
\begin{eqnarray}
\sigma^{\gamma}=  \sigma^{\gamma}_1 + \sigma^{\gamma}_2,
\end{eqnarray}
where $\sigma^{\gamma}_{1,2}$ have the similar factorized  structure
based on the factorization of soft-photon bremsstrahlung:
\begin{eqnarray}
\sigma^{\gamma}_1= \frac{\alpha}{\pi} {\rm Re} \bigl[ -\delta_1^{\lambda} +R_1 \bigr] \sigma_0, \ \ \
\label{38a}
\sigma^{\gamma}_2= \frac{\alpha}{\pi} {\rm Re} \bigl[ (-\delta_1^{\lambda} +R_1 )^* \hat \sigma_1 \bigr],
\end{eqnarray}
where
\begin{equation}
R_1=-4B \log\frac{\sqrt{s}}{2\omega} -\log^2\frac{s}{em^2}+1-\frac{\pi^2}{3} +\log^2\frac{u}{t}.
\end{equation}
The first part of the soft-photon cross section,  $\sigma^{\gamma}_1$,  cancels the IRD at the one-loop order,
while the second part, $\sigma^{\gamma}_2$, cancels the IRD a the two-loop order,
with half of $\sigma^{\gamma}_2$ going to the cancellation of IRD in the Q-part and the
other half going to treat IRD in the T-part:
\begin{eqnarray}
\sigma^{\gamma}_Q= \sigma^{\gamma}_T= \frac{1}{2}\sigma^{\gamma}_2.
\end{eqnarray}

To obtain the term $-\delta_1^{\lambda} +R_1$ in Eq. (\ref{38a}), we must calculate  the 3-dimensional
integral over the phase space of one real soft photon.
It can be done according to \cite{HooftVeltman}  in c.m.s:
\begin{eqnarray}
 -\delta_1^{\lambda} + R_1  = L(\lambda,\omega) =
-\frac{1}{4\pi}
\int\limits_{k_0<\omega}  \frac{d^3 k}{k_0}
T^{\beta}(k) T_{\beta}(k),
\label{1}
\end{eqnarray}
where
\begin{eqnarray}
T^{\alpha}(k)
=\frac{k_1^{\alpha}}{k_1k} - \frac{k_2^{\alpha}}{k_2k} +\frac{p_1^{\alpha}}{p_1k} - \frac{p_2^{\alpha}}{p_2k}.
\end{eqnarray}

The difference between the estimation relying on the soft part only and the result obtained by separation into the
soft and hard parts at lowest order is rather small (see \cite{ABIZ-prd}),
so we believe that
the soft cross section will provide the sufficient accuracy at second order as well.

At last, the cross section induced by the emission of two soft photons with a total energy less then $\omega$
can be written as:
\begin{equation}
\sigma^{\gamma\gamma}=
\frac{1}{2}
{\Bigl( \frac{\alpha}{\pi} \Bigr)}^2
\biggl( \biggl| -\delta_1^{\lambda} + R_1 \biggr|^2 - R_2 \biggr)
\sigma_0,
\end{equation}
where $\frac{1}{2}$ is a statistical factor caused by the indistinguishability of two final photons
and $R_2 = \frac{8}{3}\pi^2 |B|^2$.
The detailed calculations of $\sigma^{\gamma\gamma}$ are shown in Appendix A.

Just like $\sigma^{\gamma}$, the cross section $\sigma^{\gamma\gamma}$
is divided into equal halves, with
a half going to cancel the IRD in the Q-part and a half going to the T-part:
\begin{eqnarray}
\sigma^{\gamma\gamma}_Q= \sigma^{\gamma\gamma}_T= \frac{1}{2}\sigma^{\gamma\gamma}.
\end{eqnarray}

Combining all the terms together, we get the infrared-finite result at both the
first and second orders.
The first(second) order is given by the first(second) term on the LHS of the equation below:
\begin{equation}
{\rm Re}[\sigma_1 + \sigma^{\gamma}_1]  + (\sigma_Q + \sigma^{\gamma}_Q + \sigma^{\gamma\gamma}_Q )=
\frac{\alpha}{\pi} {\rm Re}[R_1 + \delta_1^f] \sigma_0 +
{\Bigl( \frac{\alpha}{\pi} \Bigr)}^2
{\rm Re} [\frac{1}{2} R_1^* \delta_1^f + \delta_Q^f +\frac{1}{4}R_1^*R_1 - \frac{1}{4}R_2]  \sigma_0.
\label{sum}
\end{equation}

\section{Numerical results}

For the numerical calculations
we use
$\alpha=1/137.035 999$,\
$m_W=80.398\ \mbox{GeV}$,\  and
$m_Z=91.1876\ \mbox{GeV}$ as input parameters in accordance with \cite{PDG08}.
The electron, muon, and $\tau$-lepton
masses are taken to be $m_e=0.510 998 910\ \mbox{MeV}$,  $m_\mu=0.105 658 367\ \mbox{GeV}$, $m_\tau=1.776 84\ \mbox{GeV}$,
while the quark masses for vector boson self-energy loop contributions are taken to be
$m_u=0.069 83\ \mbox{GeV}$,\ $m_c=1.2\ \mbox{GeV}$,\ $m_t=174\ \mbox{GeV}$,
$m_d=0.069 84\ \mbox{GeV}$,\ $m_s=0.15\ \mbox{GeV}$, and $m_b=4.6\ \mbox{GeV}$.
The values of the light quark masses were extracted using the fact that they provide shifts in the 
fine structure constant due to hadronic
vacuum polarization $\Delta \alpha_{had}^{(5)}(m_Z^2)$=0.02757 \cite{jeger},
where
\begin{equation}
\Delta \alpha_{had}^{(5)}(s)=\frac{\alpha}{3\pi} \sum_{f=u,d,s} Q_f^2 \biggl(\log\frac{s}{m_f^2}-\frac{5}{3}\biggr),
\label{hvpolq}
\end{equation}
and $Q_f$ is the electric charge of fermion $f$ in proton charge units $q\ (q=\sqrt{4\pi \alpha})$.

On the other hand, the contribution of hadronic vacuum polarization to the fine structure constant also can be evaluated using the 
dispersion relation:
\begin{equation}
    \Delta \alpha_{had}^{(5)}(s) =
    -\frac{s}{4\pi^2 \alpha} {\cal P}
    \int\limits_{2M_\pi^2}^\infty
    ds'
    \frac{\sigma_h(s')}{s'-s-i 0},
\end{equation}
where ${\cal P}$ means that the principle value of the integral should be considered and
$\sigma_h(s)$ is the cross section of hadron production in $e^+e^-$ annihilation.
In the case of small energies this cross section can be approximated by the cross section of the pion production channel
$e^+e^- \to \pi^+ \pi^-$:
\begin{equation}
    \sigma_h(s) = \frac{\pi\alpha^2}{3 s} \beta_\pi^3,
    \qquad
    \beta_\pi = \sqrt{1-\frac{4 M_\pi^2}{s}},
\end{equation}
thus giving the following contribution to $\Delta \alpha_{had}^{(5)}(s)$:
\begin{equation}
    \Delta \alpha_{had}^{(5)}(s) =
    \frac{\alpha}{\pi}
    \Bigl(
        \frac{1}{12} \log\Bigl(\frac{1+\beta_\pi}{1-\beta_\pi}\Bigr)
        -\frac{2}{3} - 2\beta_\pi^2
    \Bigr).
\label{hvpolp}
\end{equation}

Using Eq.~(\ref{hvpolq}) and Eq.~(\ref{hvpolp}) we can incorporate the use of the light quark masses
as parameters regulated by the hadronic vacuum polarization in our calculations. 

Finally, for the mass of the Higgs boson, we take $m_H=115\ \mbox{GeV}$.
Although this mass is still to be determined experimentally, the dependence of EWC on $m_H$ is rather weak. For the maximum soft photon energy we use $\omega = 0.05\sqrt{s}$, according to   \cite{ABIZ-prd} and \cite{5-DePo}.

Let us define the relative corrections to the Born cross section due to a specific type of contributions (labeled by $\rm C$) as
$$\delta^{\rm C} = (\sigma^{\rm C}-\sigma^0)/\sigma^0,\ \rm \ C=\mbox{1-loop}, Q, T,...  $$
The parity-violating asymmetry is defined in a traditional way,
\begin{equation}
A_{LR} =
 \frac{\sigma_{LL}+\sigma_{LR}-\sigma_{RL}-\sigma_{RR}}
      {\sigma_{LL}+\sigma_{LR}+\sigma_{RL}+\sigma_{RR}}
 =
 \frac{\sigma_{LL}-\sigma_{RR}}
      {\sigma_{LL}+2\sigma_{LR}+\sigma_{RR}},
\label{A}
\end{equation}
and the relative correction to the Born asymmetry  due to  $\rm C$-contribution
is defined as
$$\delta^{\rm C}_A = (A_{LR}^{\rm C}-A_{LR}^0)/A_{LR}^0. $$

\begin{figure}
\vspace{30mm}
\begin{tabular}{cc}
\begin{picture}(60,60)
\put(-190,-60){
\epsfxsize=9cm
\epsfysize=9cm
\epsfbox{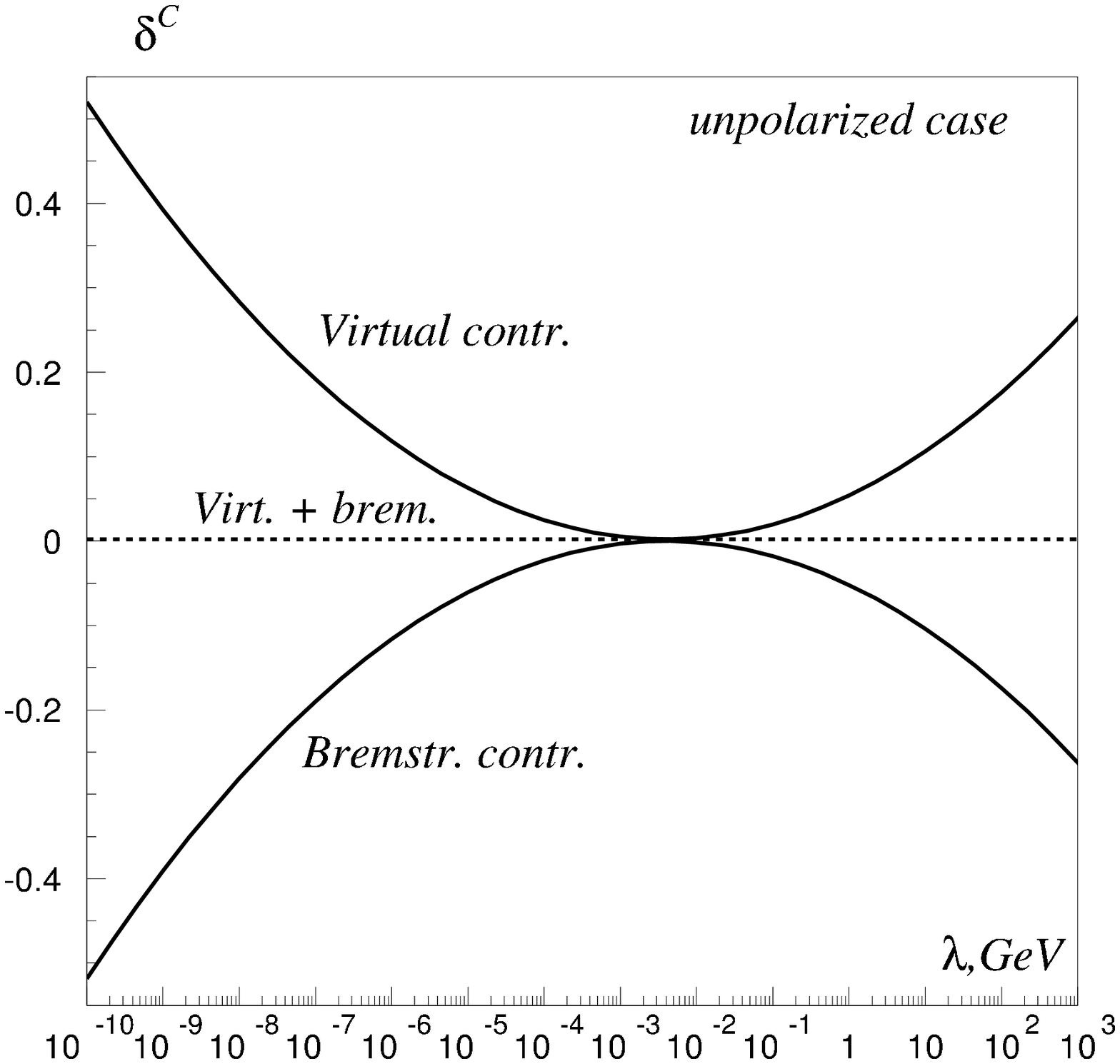}  }
\end{picture}
&
\vspace{10mm}
\begin{picture}(60,100)
\put(0,-60){
\epsfxsize=9cm
\epsfysize=9cm
\epsfbox{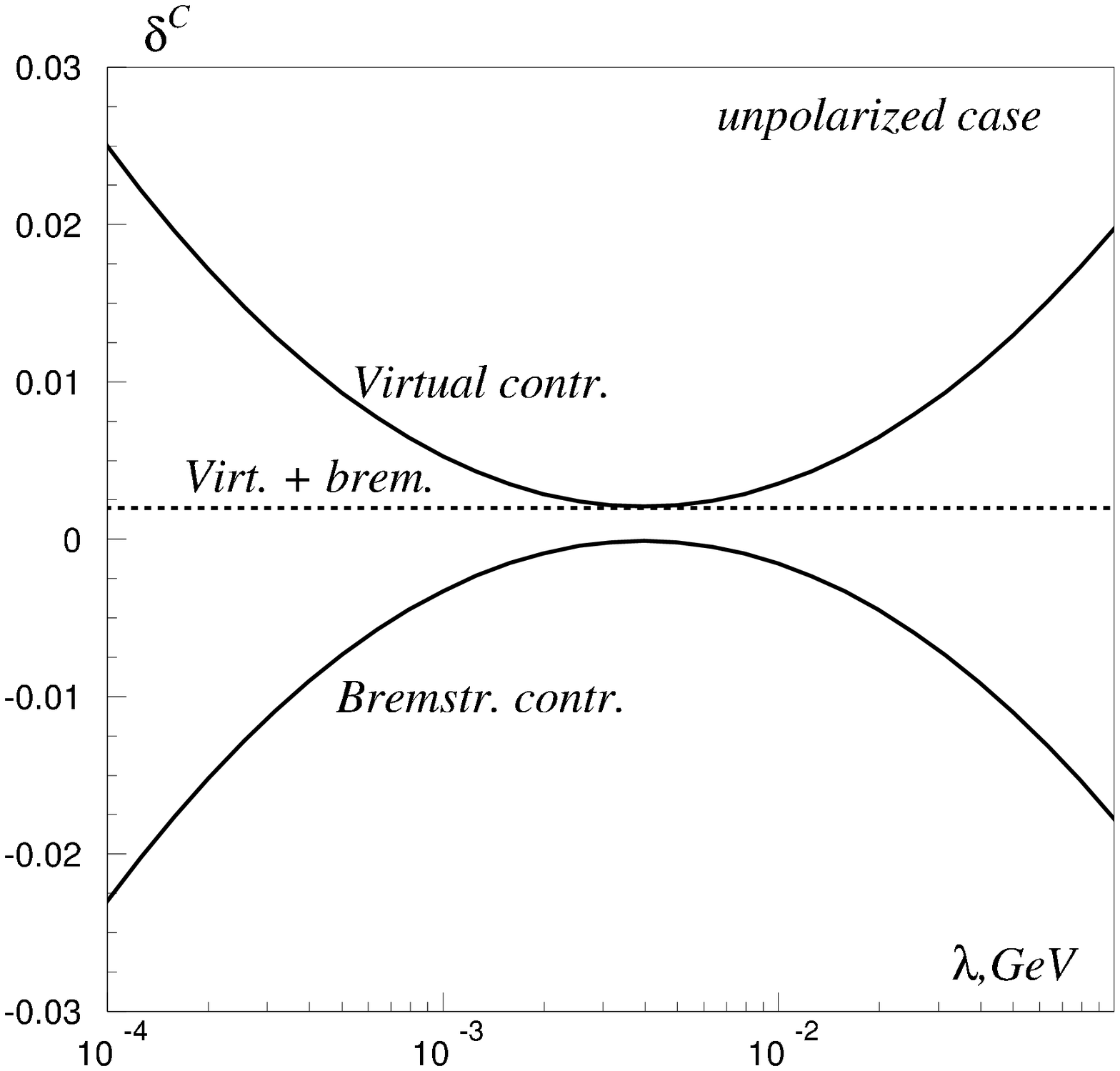}  }
\end{picture}
\end{tabular}
\vspace{10mm}
\caption{\protect\it
Virtual and bremstrahlung contributions to the relative correction to
unpolarized cross section vs. the photon mass $\lambda$ at $\theta=90^o$ and $E_{\rm lab}$ = 11 GeV.
 }
\label{f1}
\vspace{5mm}
\end{figure}

Fig.~\ref{f1}, plotted for  $\theta=90^o$ and $E_{\rm lab}$ = 11 GeV,
clearly demonstrates that the relative correction to the unpolarized cross
section is independent of the photon mass $\lambda$. We can also see
a quadratic dependence on the log scale of $\lambda$ for both the
virtual and bremstrahlung contributions.

\begin{figure}
\vspace{30mm}
\begin{tabular}{cc}
\begin{picture}(60,60)
\put(-190,-60){
\epsfxsize=9cm
\epsfysize=9cm
\epsfbox{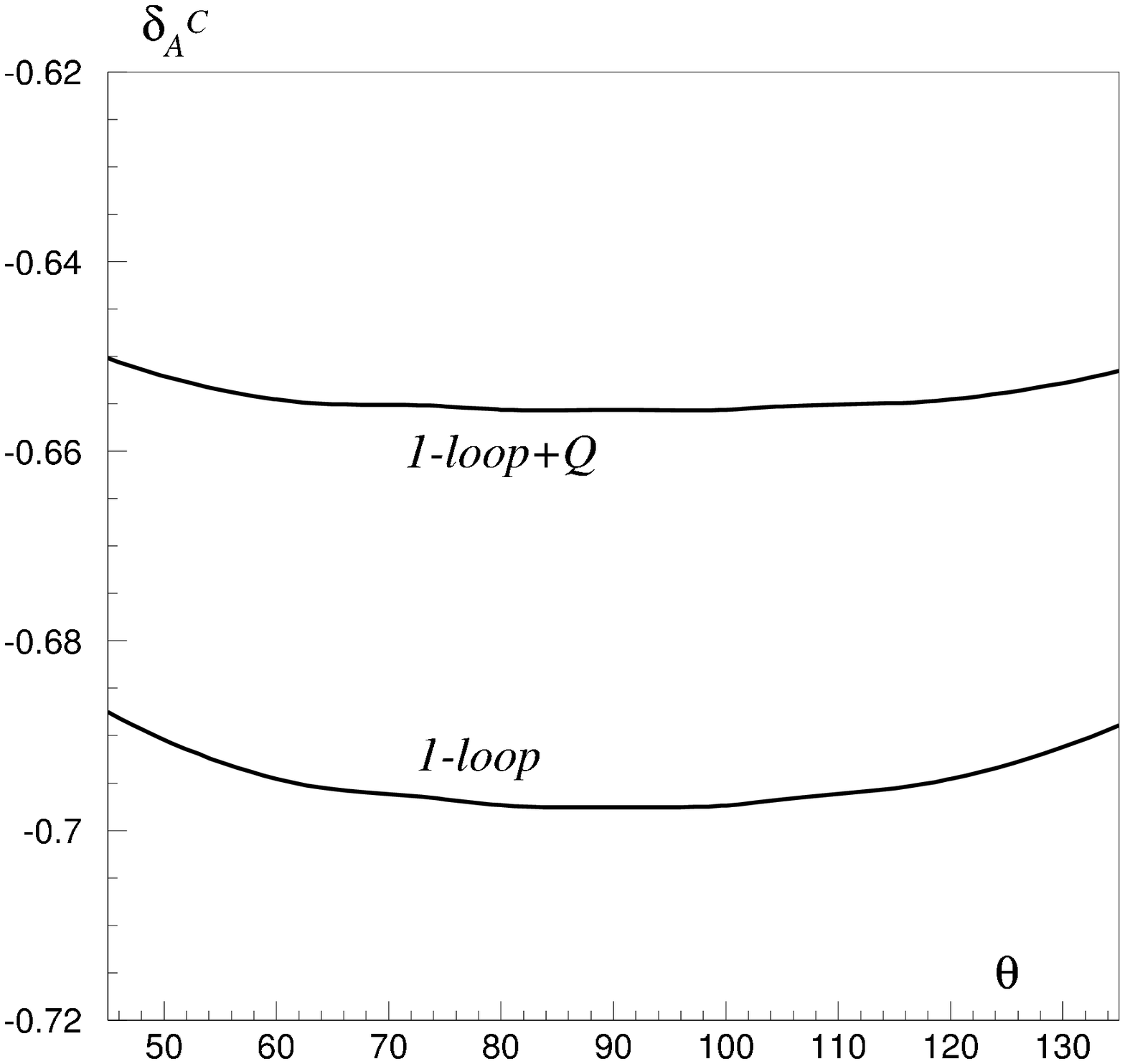}  }
\end{picture}
&
\vspace{10mm}
\begin{picture}(60,100)
\put(0,-60){
\epsfxsize=9cm
\epsfysize=9cm
\epsfbox{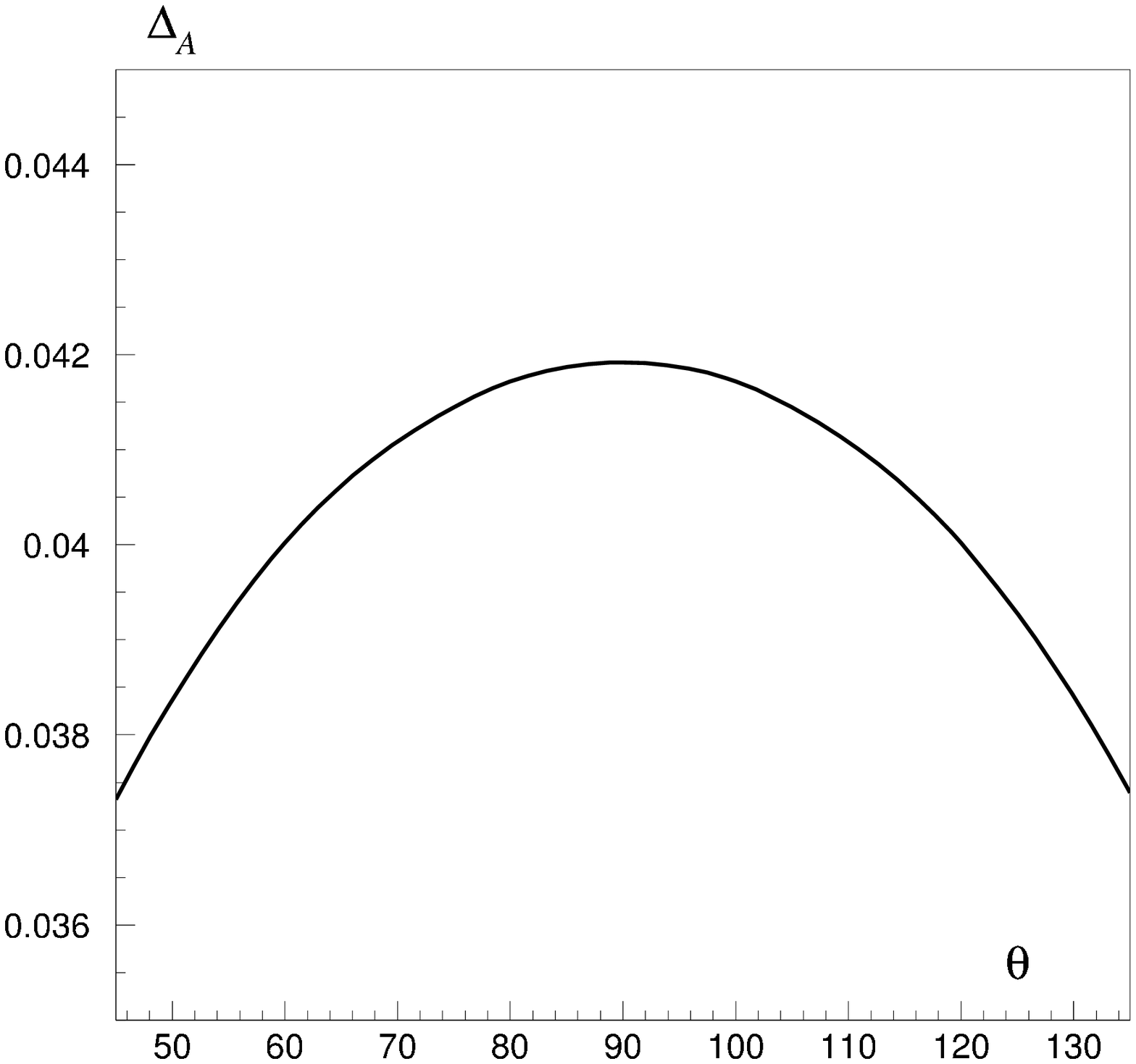}  }
\end{picture}
\end{tabular}
\vspace{10mm}
\caption{\protect\it
The relative corrections to  the asymmetry (left)
and the absolute correction $\Delta_A$ (right)
vs scattering angle $\theta$.
 }
\label{f2}
\vspace{5mm}
\end{figure}

The left frame of Fig.~\ref{f2} depicts
the relative corrections to the asymmetry at  $E_{\rm lab}$~=~11~GeV versus the
scattering angle $\theta$ in c.m.s.
The lower line shows  the  corrections to the asymmetry with only one-loop EWC taken
into account, and the upper line shows the combined  one-loop and Q-part corrections.
As expected, both of them are symmetric along the line $\theta =\pi/2$,
have a minimum at $\theta=90^o$, and depend on the scattering angle quite weakly.

The difference of these two effects is an absolute correction defined by
$$\Delta_A = (A_{LR}^{\rm 1-loop+Q}-A_{LR}^0)/A_{LR}^0 - (A_{LR}^{\rm 1-loop}-A_{LR}^0)/A_{LR}^0
= (A_{LR}^{\rm 1-loop+Q} - A_{LR}^{\rm 1-loop})/A_{LR}^0 $$
and depicted in the right frame of Fig.~\ref{f2}.
Here we can see that the Q-part gives quite a significant contribution, with $\Delta_A$ reaching a maximum of $0.0419$ at $\theta=90^o$.
Taking into account that MOLLER's planned experimental error to the PV asymmetry is $\sim 2\%$ or less,
we see that it is necessary to continue to work on the two-loop EWC, staring from the T-part.

\begin{figure}
\vspace{60mm}
\begin{tabular}{cc}
\begin{picture}(60,60)
\put(-112,-60){
\epsfxsize=10cm
\epsfysize=10cm
\epsfbox{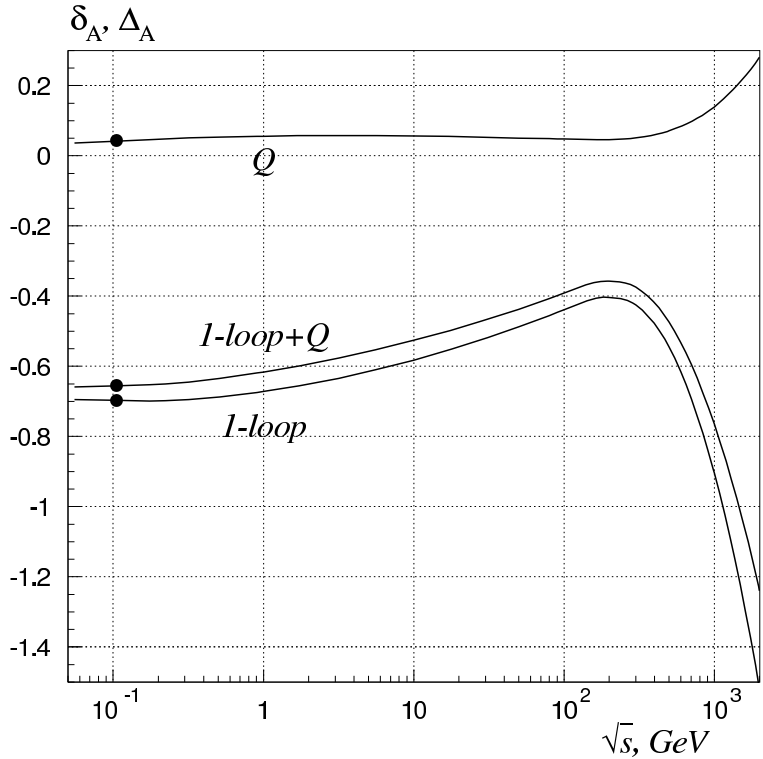}  }
\end{picture}
\end{tabular}
\vspace{20mm}
\caption{\protect\it
Relative (labeled by 1-loop+Q and 1-loop) and absolute (labeled by Q) corrections to PV-asymmetry vs $\sqrt{s}$. The filled circle corresponds to our predictions for the MOLLER experiment.}
\label{f3}
\vspace{5mm}
\end{figure}

Fig.~\ref{f3} shows the relative (labeled as 1-loop and  1-loop+Q)
corrections and absolute $\Delta_A$ corrections (labeled by Q)
versus $\sqrt{s}$ at $\theta$ = 90$^o$.
In the high-energy region ($\sqrt{s} \geq $ 50 GeV) our one-loop result (see \cite{ABIZ-prd}) is in excellent
agreement with the result from \cite{5-DePo} if we use the same set of Standard Model parameters.
As one can see from Fig.~\ref{f3}, the scale of the Q-part contribution in the low-energy region is approximately constant,
but grows sharply at $\sqrt{s} \geq m_Z$, where the weak contribution becomes
comparable to the electromagnetic.
This increasing importance of the two-loop contribution at higher energies may have a significant effect on the asymmetry measured at future $e^-e^-$-colliders.

\section{Conclusions}

Experimental investigation of M{\o}ller scattering is not only one of the oldest tools of modern physics, but also a powerful probe of new physics effects.
The new ultra-precise measurement of the weak mixing angle via 11 GeV M{\o}ller scattering planned at JLab (MOLLER) -- as well as experiments proposed at future high-energy electron colliders -- will require that the higher-order effects to be taken into account with the highest precision possible.

In this work, we build on the study of the one-loop electroweak radiative corrections to the cross-section asymmetry of the polarized M{\o}ller scattering at 11 GeV initiated by our group in \cite{ABIZ-prd}, and address some of the two-loop effects.
At this stage, we perform a detailed calculation for the part of the two-loop electroweak radiative correction induced by squaring one-loop diagrams.

The two-loop EWC to the Born ($\sim M_0M_0^+$) cross section is divided
into the T-part, which is the interference of Born and two-loop diagrams ($ \sim 2 \mbox{\rm Re} M_{0} M_{2-loop}^+$), and the Q-part, induced by quadratic one-loop amplitudes ($ \sim M_1M_1^+$), which we evaluate here. The results are presented in both numerical and analytical form, with the infrared divergence explicitly cancelled. Also, we clearly demonstrate the important role of the imaginary part of amplitude, which is consistently taken into consideration both in the infrared-finite and divergent terms.

As one can see from our numerical data, at the MOLLER kinematic conditions, the part of the
NNLO EWC we considered in this work can increase the asymmetry by up to $\sim 4$\%.
The corrections depend quite significantly on the energy and scattering angles; at the high-energy region of $\sqrt{s} \sim 1000$~GeV achievable in the planned experimental program of the ILC, the estimated contribution of the quadratic EWC can reach +14\%; for 3 TeV at CLIC, it would be +42\%.
We see that the large size of the Q-part demands detailed and consistent consideration of the T-part, which will be the next task of our group.
It is impossible to say at this time if the Q-part will be partially enhanced or cancelled by
other two-loop radiative corrections, although it seems probable that the two-loop EWC may be larger
than previously thought. Although an argument can be made that the two-loop corrections are suppressed
by a factor of $\alpha\pi$ relative to the one-loop corrections (see \cite{Petr2003}, for example),
we are reluctant to conclude that they can be dismissed,  especially in the  light of 2\% uncertainty
to the asymmetry promised by MOLLER.

Since the problem of EWC for the M{\o}ller scattering asymmetry is rather involved,
a tuned step-by-step comparison between different calculation approaches is essential.
One of the important results of this work is the correctness of our calculations,
which was controlled by a comparison of the results obtained from the equations derived
by hand with the numerical data obtained by a semi-automatic approach based on FeynArts,
FormCalc, LoopTools and Form. These base languages have already been successfully employed
in similar projects (\cite{ABIZ-prd}, \cite{arx-2}), so we are highly confident in their reliability.

In the future, we plan to address the remaining two-loop electroweak corrections which
may be required by the promised experimental precision of the MOLLER experiment and
experiments planned at ILC.

\ \\
\ \\

\section{ACKNOWLEDGMENTS}

We are grateful to Y. Bystritskiy and T. Hahn
for stimulating discussions.
A. A. and S. B. thank the Theory Center at Jefferson Lab, and V. Z. thanks
Acadia University for hospitality in 2011. This work was supported by the
Natural Sciences and Engineering Research Council of Canada
and Belarus scientific program "Convergence".

\ \\
\ \\


\newpage

\appendix

\section{Detailes of calculations for the case of emission of two real soft photons}
\label{sec:appendix-a}

\begin{figure}
\vspace{10mm}
\begin{tabular}{cc}
\begin{picture}(60,60)
\put(-150,-60){
\epsfxsize=7cm
\epsfysize=7cm
\epsfbox{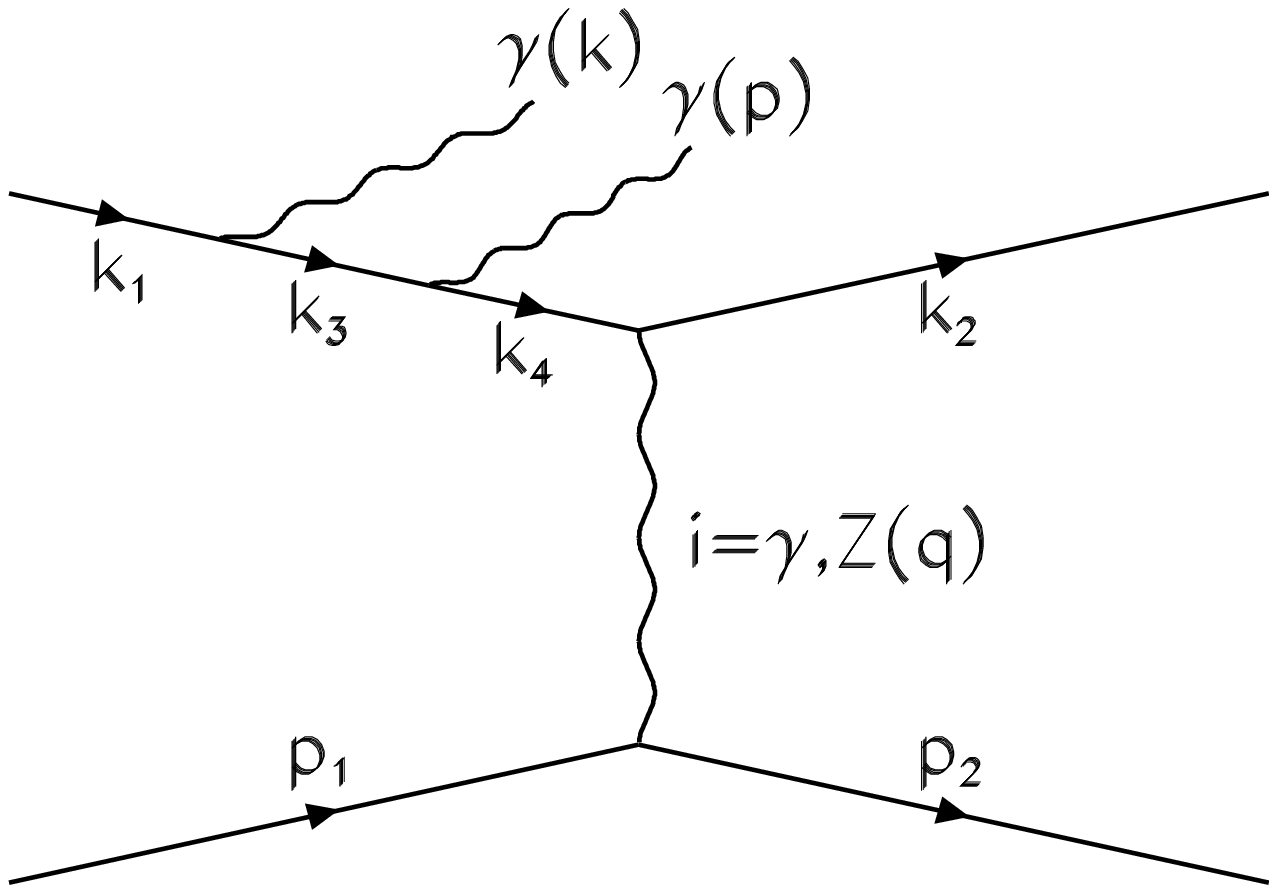}  }
\end{picture}
&
\begin{picture}(60,100)
\put(-20,-60){
\epsfxsize=7cm
\epsfysize=7cm
\epsfbox{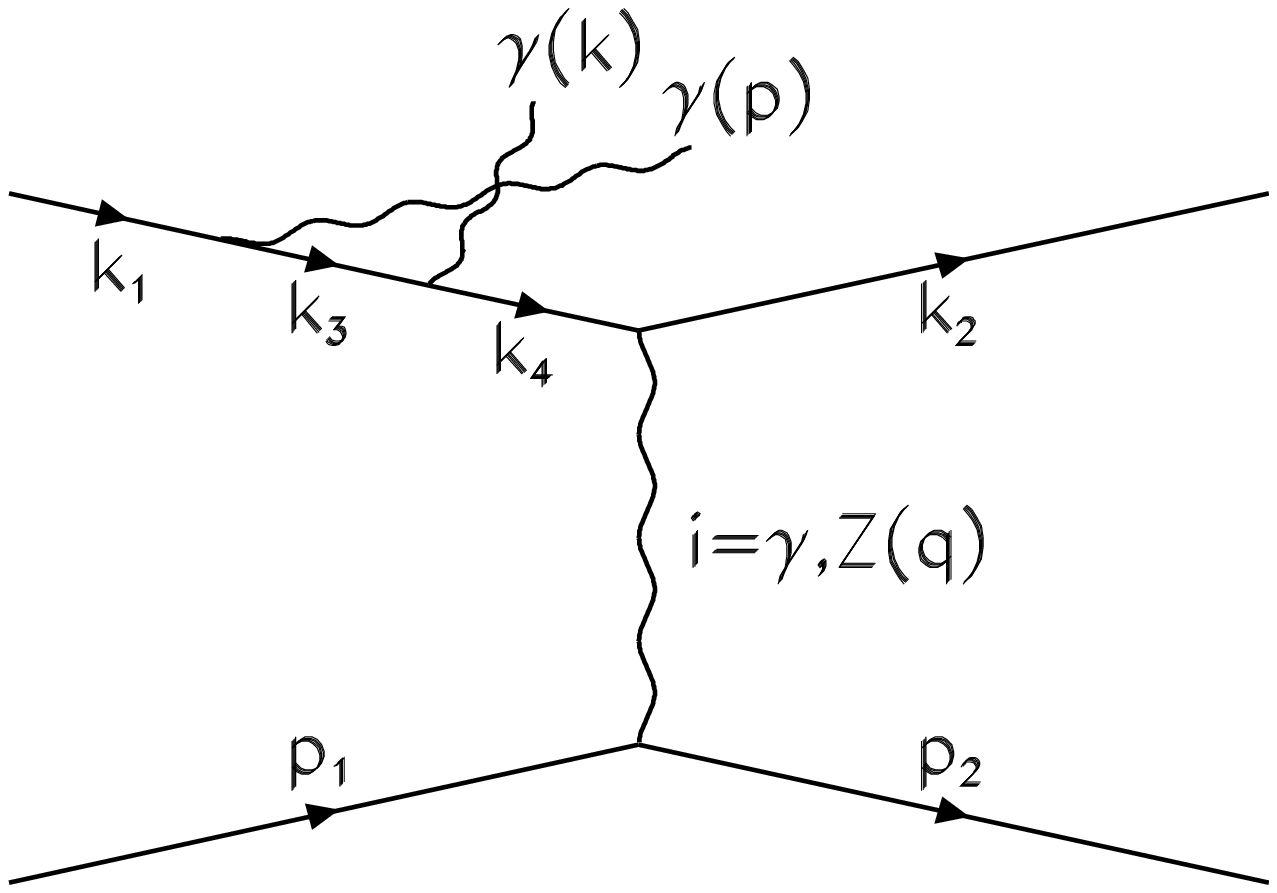}  }
\end{picture}
\end{tabular}
\vspace{10mm}
\caption{\protect\it
Double-bremsstrahlung diagrams for M{\o}ller scattering in $t$-channel
corresponding to $M_{11}^i$.
The  u-channel diagrams are obtained by interchange $k_2 \leftrightarrow p_2$. }
\label{br11r}
\vspace{5mm}
\end{figure}

First, let us calculate the amplitudes corresponding to the emission
of two real soft photons (see Fig.~\ref{br11r}),
\begin{equation}
e^-(k_1,\xi)+e^-(p_1,\eta) \rightarrow e^-(k_2)+e^-(p_2)+\gamma(k)+\gamma(p)
\label{00}
\end{equation}
in $t$- and $u$-channels
with $i$-boson exchange ($i=\gamma,\ Z$).
The amplitudes are labeled as $M_{11}^i,\ M_{12}^i,\ M_{13}^i,... $,
where the first (second) subscript denotes the origin of the first $\gamma(k)$ (second $\gamma(p)$) emitted photon:
1 -- emitted from electron $e^-(k_1)$,
2 -- from electron $e^-(k_2)$,
3 -- from electron $e^-(p_1)$ and
4 -- from electron $e^-(p_2)$.
The exact expression for $M_{11}^i$ is the following:
\begin{eqnarray}
M_{11}^i = &&
i (2\pi e)^4 N_{k_1} N_{k_2} N_{p_1} N_{p_2} N_{p} N_{k} \cdot
\bar u(k_2) \gamma^{\mu} (v^i-a^i\gamma_5)
\frac{1}{\hat k_4-m} \gamma^{\alpha} e_{\alpha}(p)
\frac{1}{\hat k_3-m} \gamma^{\beta} e_{\beta}(k) u(k_1) \cdot
\nonumber \\ &&
\bar v(p_2) \gamma_{\mu} (v^i-a^i\gamma_5) v(p_1) \cdot \frac{1}{q^2-m_i^2} \cdot
\delta(k_1+p_1-k_2-p_2-k-p),
\label{M11}
\end{eqnarray}
where $N_k=\frac{1}{(2\pi)^{3/2}}\frac{1}{\sqrt{2k_0}}$.
Using the Dirac equation and taking $k\rightarrow 0$, we can simplify
\begin{eqnarray}
\frac{1}{\hat k_3-m} \gamma^{\beta} u(k_1)
&&
=\frac{\hat k_1 -\hat k+m}{(k_1-k)^2-m^2} \gamma^{\beta} u(k_1)
\approx \frac{\hat k_1 +m}{-2k_1k} \gamma^{\beta} u(k_1) =
\nonumber \\ &&
= \frac{1}{-2k_1k} (2k_1^{\beta} +\gamma^{\beta}[-\hat k_1+m]) u(k_1)
= -\frac{k_1^{\beta}}{k_1k} u(k_1).
\label{simmpp}
\end{eqnarray}
Analogously, at $k, p \rightarrow 0$,
\begin{eqnarray}
\frac{1}{\hat k_4-m} \gamma^{\alpha} u(k_1)
= \frac{k_1^{\alpha}}{-k_1(k+p)+kp} u(k_1).
\label{simmpp2}
\end{eqnarray}

Finally, the amplitude $M_{11}^i$ at $k, p \rightarrow 0$
has the following form, with the convenient factorization from the Born amplitude:
\begin{eqnarray}
M_{11}^i |_{k, p \rightarrow 0} = &&
e^2 N_{p} N_{k} \cdot
e_{\alpha}(p) e_{\beta}(k) \cdot
\frac{k_1^{\alpha} k_1^{\beta}}{(-k_1k)(-k_1(k+p)+kp)} \cdot M_{0}^i.
\label{M110}
\end{eqnarray}

In the same manner, we get
\begin{eqnarray}
M_{22}^i |_{k, p \rightarrow 0} = &&
e^2 N_{p} N_{k} \cdot
e_{\alpha}(p) e_{\beta}(k) \cdot
\frac{k_2^{\alpha} k_2^{\beta}}{(k_2p)(k_2(k+p)+kp)} \cdot M_{0}^i,
\nonumber \\
M_{12}^i |_{k, p \rightarrow 0} = &&
e^2 N_{p} N_{k} \cdot
e_{\alpha}(p) e_{\beta}(k) \cdot
\frac{k_2^{\alpha} k_1^{\beta}}{(k_2p)(-k_1k)} \cdot M_{0}^i,
\nonumber \\
M_{33}^i |_{k, p \rightarrow 0} = &&
e^2 N_{p} N_{k} \cdot
e_{\alpha}(p) e_{\beta}(k) \cdot
\frac{p_1^{\alpha} p_1^{\beta}}{(-p_1k)(-p_1(k+p)+kp)} \cdot M_{0}^i,
\nonumber \\
M_{44}^i |_{k, p \rightarrow 0} = &&
e^2 N_{p} N_{k} \cdot
e_{\alpha}(p) e_{\beta}(k) \cdot
\frac{p_2^{\alpha} p_2^{\beta}}{(p_2p)(p_2(k+p)+kp)} \cdot M_{0}^i,
\nonumber \\
M_{34}^i |_{k, p \rightarrow 0} = &&
e^2 N_{p} N_{k} \cdot
e_{\alpha}(p) e_{\beta}(k) \cdot
\frac{p_2^{\alpha} p_1^{\beta}}{(p_2p)(-p_1k)} \cdot M_{0}^i,
\nonumber \\
M_{13}^i |_{k, p \rightarrow 0} = &&
e^2 N_{p} N_{k} \cdot
e_{\alpha}(p) e_{\beta}(k) \cdot
\frac{k_1^{\alpha} p_1^{\beta}}{(-k_1p)(-p_1k)} \cdot M_{0}^i,
\nonumber \\
M_{14}^i |_{k, p \rightarrow 0} = &&
e^2 N_{p} N_{k} \cdot
e_{\alpha}(p) e_{\beta}(k) \cdot
\frac{k_1^{\alpha} p_2^{\beta}}{(-k_1p)(p_2k)} \cdot M_{0}^i,
\nonumber \\
M_{23}^i |_{k, p \rightarrow 0} = &&
e^2 N_{p} N_{k} \cdot
e_{\alpha}(p) e_{\beta}(k) \cdot
\frac{k_2^{\alpha} p_1^{\beta}}{(k_2p)(-p_1k)} \cdot M_{0}^i,
\nonumber \\
M_{24}^i |_{k, p \rightarrow 0} = &&
e^2 N_{p} N_{k} \cdot
e_{\alpha}(p) e_{\beta}(k) \cdot
\frac{k_2^{\alpha} p_2^{\beta}}{(k_2p)(p_2k)} \cdot M_{0}^i.
\label{allM0}
\end{eqnarray}

Now we need to sum the terms generated by the substitution $k \leftrightarrow p$.
For the 11--, 22--, 33--, 44--cases it works as the following:
\begin{eqnarray}
M_{11}^i |_{k, p \rightarrow 0} + (k \leftrightarrow p) =
&&
e^2 N_{p} N_{k}  M_{0}^i
\Bigl(
\frac{e_{\alpha}(p) e_{\beta}(k) k_1^{\alpha} k_1^{\beta}}{(-k_1k)(-k_1(k+p)+kp)}
+\frac{e_{\alpha}(k) e_{\beta}(p) k_1^{\alpha} k_1^{\beta}}{(-k_1p)(-k_1(k+p)+kp)}
\Bigr)
\nonumber \\
&&
=e^2 N_{p} N_{k}  M_{0}^i
\Bigl(
\frac{1}{-k_1k} + \frac{1}{-k_1p}
\Bigr)
\frac{e_{\alpha}(p) e_{\beta}(k) k_1^{\alpha} k_1^{\beta}}{(-k_1(k+p)+kp)}
\nonumber \\
&&
\approx e^2 N_{p} N_{k}  M_{0}^i
\frac{e_{\alpha}(p) e_{\beta}(k) k_1^{\alpha} k_1^{\beta}}{(k_1k) (k_1p)}.
\label{M110kp}
\end{eqnarray}

As a result, the total $t(u)$-channel amplitude is given by
\begin{eqnarray}
M_{t(u)}^i |_{k, p \rightarrow 0} =
e^2 N_{p} N_{k} \cdot
e_{\alpha}(p) e_{\beta}(k) \cdot T^{\alpha}(p) T^{\beta}(k)  \cdot  M_{0,t(u)}^i.
\label{Mtu}
\end{eqnarray}
Then, the cross section of (\ref{00}) has the form
\begin{eqnarray}
\sigma^{\gamma\gamma}
=&&
\sigma_0
\frac{1}{2}
\int\limits_{k_0+p_0<\omega}  d^3 k d^3 p \cdot
(eN_k)^2 (eN_p)^2
T^{\alpha}(p) T_{\alpha}(p)
T^{\beta}(k) T_{\beta}(k)
\nonumber \\
&&
=
\sigma_0
{\Bigl( \frac{\alpha}{\pi} \Bigr)}^2
\frac{1}{2}
\Bigl( \frac{1}{4\pi} \Bigr)^2
\int\limits_{k_0+p_0<\omega}  \frac{d^3 k}{k_0} \frac{d^3 p}{p_0} \cdot
T^{\alpha}(p) T_{\alpha}(p)
T^{\beta}(k) T_{\beta}(k).
\label{fin}
\end{eqnarray}

It is possible to prove that
\begin{eqnarray}
\Bigl( \frac{1}{4\pi} \Bigr)^2
\int\limits_{k_0+p_0<\omega}  \frac{d^3 k}{k_0} \frac{d^3 p}{p_0} \cdot
T^{\alpha}(p) T_{\alpha}(p)
T^{\beta}(k) T_{\beta}(k) =
  |L(\lambda,\omega)|^2 - R_2.
\label{s2}
\end{eqnarray}
If we change the condition $k_0+p_0<\omega$ to simply
 $k_0<\omega,\ p_0<\omega$, the term $R_2$ will go to zero. However, the conditions  $k_0<\omega,\ p_0<\omega$ are not valid from the experiment point of view.

Let us  calculate $R_2$ exactly. First, we introduce the notation
\begin{eqnarray}
I(a_1,a_2;b_1,b_2) =
\Bigl( \frac{1}{4\pi} \Bigr)^2
\int\limits_{a_1<k_0<a_2,\ b_1<p_0<b_2}  \frac{d^3 k}{k_0} \frac{d^3 p}{p_0} \cdot
T^{\alpha}(p) T_{\alpha}(p)
T^{\beta}(k) T_{\beta}(k).
\end{eqnarray}
From Eq.(\ref{1}) it is obvious that
\begin{eqnarray}
I(0,a;0,b) = L(\lambda,a) L(\lambda,b)^{*},
\label{2to1}
\end{eqnarray}
which, of cause, works at $\lambda \ll a,b$ only.
At $da,db \ll a,b$, the simple geometry considerations give the equation
$$I(a,a+da;b,b+db) =  I(0,a+da;0,b+db)+I(0,a;0,b)-I(0,a;0,b+db)-I(0,a+da;0,b).$$
Simplifying using Eq.(\ref{2to1}) and Eq.(\ref{1}), we get
\begin{eqnarray}
I(a,a+da;b,b+db)
=  16 |B|^2 \log \frac{a+da}{a} \log \frac{b+db}{b}
\approx  16 |B|^2 \frac{da}{a} \frac{db}{b}.
\label{ii}
\end{eqnarray}

Finally, comparing Eq.(\ref{s2}) and Eq.(\ref{ii}), we conclude:
\begin{eqnarray}
R_2
=\sum\limits_{\Omega} I(a,a+da;b,b+db)
=16 |B|^2 \int\limits_0^\omega \frac{da}{a} \int\limits_{\omega-a}^\omega \frac{db}{b}
= 16 |B|^2 \mbox{Li}_2\Bigl(\frac{a}{\omega}\Bigr)\Bigr.\Bigr|_0^{\omega}
= \frac{8}{3}\pi^2 |B|^2.
\label{r2222}
\end{eqnarray}
Here, $\Omega = \{a < \omega \} \cap \{ b < \omega \} \cap \{ a+b > \omega \}$.
Our result for $R_2$ exactly agrees with \cite{KuFa}.

\end{document}